\documentclass[aps,prx,reprint,twocolumn,superscriptaddress,floatfix,nofootinbib,longbibliography]{revtex4-1}
\usepackage{amsthm}
\usepackage{amsmath,amssymb,color,comment,physics}
\usepackage[makeroom]{cancel}
\usepackage[caption=false]{subfig}
\usepackage{mathrsfs}
\usepackage{graphicx}
\usepackage{subfig}
\usepackage[countmax]{subfloat}
\usepackage[english]{babel}
\usepackage{dsfont}
\usepackage{bm}
\usepackage[bookmarks=true,colorlinks,linkcolor=OrangeRed,urlcolor=NavyBlue,citecolor=RoyalBlue]{hyperref}
\usepackage[dvipsnames]{xcolor}
\usepackage{braket}
\usepackage{mathtools}
\usepackage{soul}
\usepackage{comment}
\usepackage{booktabs}
\usepackage{makecell}
\usepackage[x11names]{xcolor}

\definecolor{mygreen}{rgb}{0.25,0.5,0.25}

\begin{document}

\preprint{APS/123-QED}

\title{Entanglement and information scrambling in long-range measurement-only circuits}

\author{Abigail McClain Gomez$^+$}
\email{amcclain@g.harvard.edu}
\affiliation{Department of Physics, Harvard University, 17 Oxford Street Cambridge, MA 02138, USA}

\author{Fiona Abney-McPeek$^+$}
\email{fabneymc@mit.edu}
\affiliation{Department of Physics, MIT, 77 Massachusetts Avenue Cambridge, MA 02139, USA}
\affiliation{Department of Physics, Harvard University, 17 Oxford Street Cambridge, MA 02138, USA}

\author{Hong-Ye Hu}
\affiliation{Department of Physics, Harvard University, 17 Oxford Street Cambridge, MA 02138, USA}

\author{Susanne F.~Yelin}
\affiliation{Department of Physics, Harvard University, 17 Oxford Street Cambridge, MA 02138, USA}

\author{Ceren B.~Da\u{g}}
\email{cbdag@iu.edu}
\affiliation{Department of Physics, Indiana University, Bloomington, Indiana 47405, USA}
\affiliation{Department of Physics, Harvard University, 17 Oxford Street Cambridge, MA 02138, USA}
\affiliation{ITAMP, Harvard-Smithsonian Center for Astrophysics, Cambridge, Massachusetts, 02138, USA}

\def\thefootnote{+}\footnotetext{These authors contributed equally to this work.}

\begin{abstract}

Measurement-only circuits provide a minimal setting in which repeated local projections can either generate or suppress many-body entanglement, giving rise to measurement-induced phase transitions and dynamical regimes, that might have no unitary counterpart. Here we investigate entanglement and information transitions in one-dimensional measurement-only Clifford circuits with long-range two-qubit parity checks. By tuning both the measurement range and density per layer, we uncover a broad set of phases whose classification requires probes beyond entanglement entropy, such as mutual information, tripartite mutual information, purification from an ancilla, and Bell-cluster statistics. We map phase diagrams using large-scale Clifford simulations for two protocols: a random-basis design in which each measurement is randomly chosen from $\lbrace XX,YY,ZZ \rbrace$, and a single-basis design in which the basis is fixed within each layer but varies between layers, hence introducing more structure to the circuit. 

We map the trajectory-averaged entanglement entropy to a 
two-dimensional statistical mechanics model by extending a replica-based method to random-basis measurement-only circuits, and show that a continuous-time limit yields an effective long-range XX hamiltonian in the steady state. This connection links the observed volume-law to sub-volume-law entanglement transition to the boundary between a continuous symmetry broken phase and a critical XY phase. Strikingly, in
structured (single-basis) circuits we find a phase in which volume-law and long-range entanglement coexists with rapid, size-independent purification of an ancilla qubit, and the absence of scrambling, highlighting measurement-only circuits as a promising route to efficiently preparing highly entangled and technologically useful quantum states.

\end{abstract}

\maketitle

\section{Introduction}

The interplay between unitary time evolution and projective measurements has opened an active frontier in quantum many-body physics in recent years, emphasizing the role of entanglement structure in quantum phases of matter and phase transitions. A particular setup of these studies is a monitored quantum circuit, in which the quantum system evolves under both coherent gates and stochastic measurements. Monitored quantum circuits can exhibit measurement-induced phase transitions (MIPTs) that separate phases of extensive (volume-law) and bounded (area-law) entanglement scaling \cite{PhysRevB.98.205136,PhysRevX.9.031009,PhysRevB.99.224307,PhysRevB.101.104301,PhysRevX.10.041020,PhysRevB.101.104302}. These transitions do not arise from symmetry breaking; rather, they emerge through the competition between entangling dynamics and the disentangling effect of projective measurements, which are visible only in trajectory-averaged observables. Not long after the discovery of the MIPTs present in hybrid unitary-projective random circuits, measurement-only circuits (MoCs) devoid of any unitary dynamics were found to generate entangled states with different properties, also leading to MIPTs \cite{PhysRevB.102.094204,PhysRevX.11.011030,nahum_entanglement_2020,PhysRevResearch.6.L042063}. The emergence of such phases is tied to the noncommutativity of the measurements \cite{lavasani_measurementinduced_2021,PhysRevResearch.3.023200,PhysRevB.106.104307}, and in some cases MoCs can be mapped to effective classical models that shed light on the underlying steady-state physics \cite{PhysRevB.102.094204, nahum_entanglement_2020}. 
\begin{figure*}[!htb]
\centering
\subfloat[][]{
\includegraphics[width=0.8\textwidth]{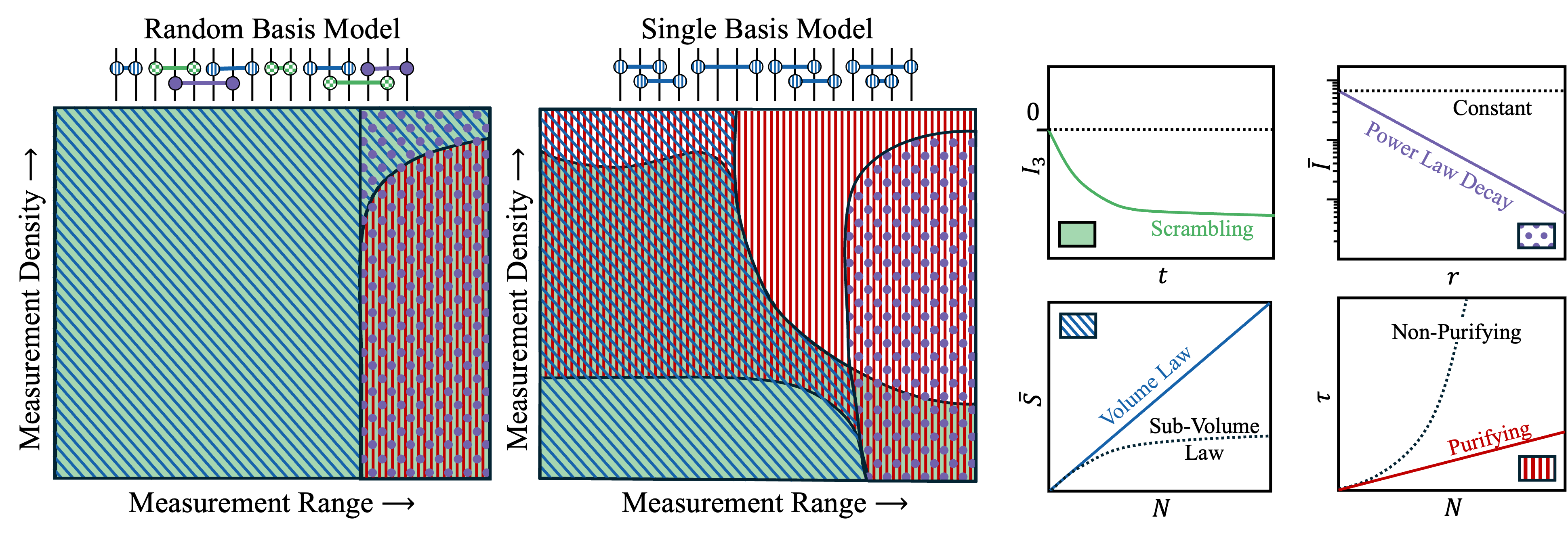}\label{subfig:phases-overview}
}

\subfloat[][]{
\includegraphics[width=0.8\textwidth]{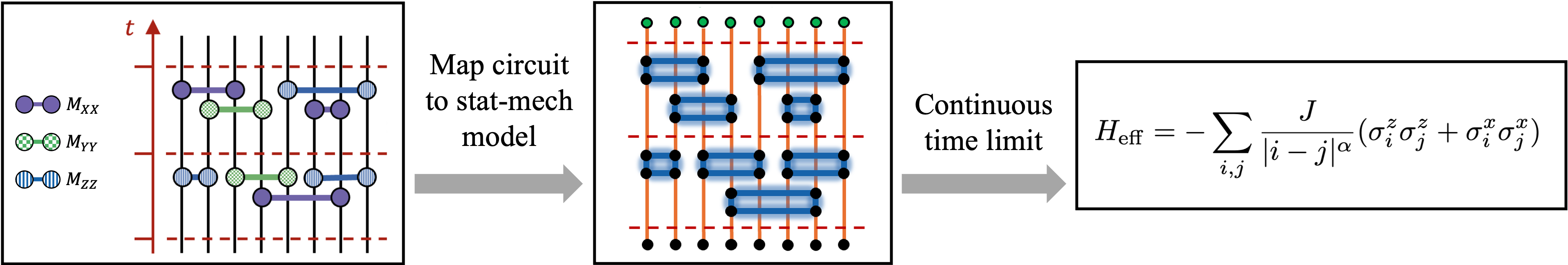}\label{subfig:statmech-overview}
}
\caption{Overview of the central results. \ref{subfig:phases-overview} The schematics of the phase diagrams for random-basis and single-basis circuit models with respect to measurement density and range obtained via Clifford circuit simulations. We obtain three and six different phases in random-basis and single-basis models, respectively. Each phase is depicted with a unique combination of scrambling, entanglement entropy, mutual information and purification properties. The markers to describe a phase are depicted in the legend plots right next to the phase diagram schematics: green, dotted-purple, blue diagonal stripes and red vertical stripes represent, respectively, a scrambling phase, power-law decaying mutual information in distance, volume-law entanglement entropy scaling in system size and a purifying phase. Not using these markers in a phase corresponds, respectively, a non-scrambling phase, constant mutual information in distance, sub-volume law entanglement entropy scaling in system size and a non-purifying phase. \ref{subfig:statmech-overview} The theoretical method applied to prove the volume-law to sub-volume law transition in random-basis model. We map the trajectory averaged entanglement entropy of measurement-only random-basis circuit to the free energy of an emergent statistical mechanics model. The middle panel specifies the lattice geometry of the resulting two-dimensional statistical mechanics model, with black dots, orange links and blue rectangles representing lattice sites, couplings and interactions, respectively. This formalism leads to an equivalence between the steady state of the circuit and the ground state of a power-law interacting XX Hamiltonian, once the continuous-time limit is taken. The volume-law and sub-volume law phases in the steady state of the circuit map to spontaneous symmetry breaking and XY phases of the effective Hamiltonian, respectively.}
\end{figure*}

Long-range interactions play a central role in quantum many-body physics. They appear naturally in both solid-state and engineered matter, such as trapped ions \cite{RevModPhys.93.025001}, Rydberg atom arrays \cite{browaeys2020many}, dipolar atom ensembles in traps \cite{chomaz2022dipolar} and optical lattices \cite{Su_2023}, dipolar spin ensembles \cite{davis2023probing}, and cavity matter hybrids \cite{RevModPhys.85.553}. In these systems, interactions that slowly decay with distance can fundamentally alter the nature of correlations, entanglement growth, and thermalization processes compared to their short-range counterparts \cite{PhysRevLett.111.207202,richerme2014non,neyenhuis2017observation,PhysRevLett.114.157201,defenu2024out}. Such phenomena can also be studied in the framework of quantum circuits with long-range unitary gates or with long-range weak measurements, which mimic nonlocal interactions \cite{PhysRevResearch.5.L012031,PRXQuantum.4.030325, PhysRevLett.128.010604,sharma2022measurement,y5r3-tv78,PhysRevLett.130.120402}. In fact, MoCs with long-range projective measurements have recently been introduced \cite{kuno_phase_2023} and claimed to exhibit both a scrambling transition from a fast scrambling to non-scrambling regime and an MIPT between volume and sub-volume law entanglement phases.

Insights into MIPTs and the entanglement structures of the steady states produced by random circuits have fueled an interest in the dynamics of information scrambling in hybrid-projective circuits \cite{PhysRevResearch.4.013174,PhysRevLett.131.220404,PRXQuantum.4.030325,PhysRevResearch.5.L012031} and in MoCs \cite{kuno_phase_2023}. Scrambling is the process where local observables spatially far from each other get correlated in time, and hence become invisible to local probes \cite{PRXQuantum.5.010201}. It has been an instrumental tool in understanding information-theoretic properties of black hole models \cite{hayden_black_2007,lashkari_fast_2013a,Landsman_2019} and strongly correlated condensed matter systems \cite{patel2017quantum,2017PhRvB..95f0201S,2017NJPh...19f3001B,iyoda_scrambling_2018,PhysRevB.98.045102,PhysRevB.98.134303,PhysRevX.8.021013,PhysRevX.8.031058,PhysRevLett.122.220601,PhysRevB.101.104415}. Of particular interest is the phenomenon of fast scrambling -- that is, scrambling that occurs in a timescale that grows logarithmically with system size, saturating the theoretical limit \cite{PhysRevA.94.040302,2016arXiv160701801Y,PhysRevLett.126.200603,2025arXiv250819075H,2025arXiv250926310S}.

In this work, measurement-only circuits are used as a controlled platform to expose how the \textit{range, density, and structure of measurements} jointly shape entanglement and information properties of many-body quantum states. The circuits are built from two-qubit parity checks drawn from \(\{X_iX_j,\,Y_iY_j,\,Z_iZ_j\}\) with long-range coupling strengths \(J_{ij}\propto |i-j|^{-\alpha}\), and are studied in two different designs, which become the same in the sparse density limit: a random-basis protocol where each check is chosen independently, and a single-basis protocol where the basis is fixed within a layer but varies between layers. Across both designs we map steady-state phase diagrams using large-scale Clifford stabilizer simulations \cite{PyCliffordRepo}, characterizing not only entanglement scaling but also mutual information, tripartite mutual information, purification from an ancilla, and Bell-cluster statistics.

Our central results are summarized in Figs.~\ref{subfig:phases-overview} and~\ref{subfig:statmech-overview}, where we schematically highlight the richness of phases emerging in the circuit designs with respect to measurement density and range. One key result is a \emph{measurement-range driven entanglement transition}. On the theory side, we develop a replica-based mapping for random-basis measurement-only circuits that relates trajectory-averaged entanglement entropies to the free energy of a two-dimensional statistical mechanics model. In the continuous-time limit this construction yields an effective long-range XX hamiltonian evolving in imaginary time, Fig.~\ref{subfig:statmech-overview}, predicting a transition from a volume-law to a sub-volume critical phase at a characteristic range near \(\alpha\sim 3\) by mapping them to a continuous symmetry broken and XY phases, respectively. This prediction is borne out by the numerics across a broad span of measurement densities, Fig.~\ref{subfig:phases-overview}. Introducing additional circuit structure via the single-basis protocol preserves the transition while shifting the transition point with measurement density, indicating that the transition is robust but its location is tunable by density. 

Beyond entanglement scaling, the combined diagnostics reveal multiple distinct regimes and separations between entanglement entropy, mutual information, scrambling, and purification (Fig.~\ref{subfig:phases-overview}). In random-basis circuits, scrambling is ubiquitous and the scrambling timescale is strongly density-controlled, crossing over from parametrically slow scrambling in sparse circuits to fast-scrambling behavior with \(\log N\) scaling in dense circuits. In contrast, the single-basis protocol exhibits clear scrambling-to-nonscrambling transitions driven by both measurement range and density, demonstrating that modest measurement structure can qualitatively reorganize information spreading even when the entanglement transition remains.

We also bias the measurement basis and introduce a projective XXZ model in the short-range and sparse measurement limit, inspired by the projective Ising model \cite{PhysRevB.102.094204}. The analysis uncovers a distinct transition between sub-volume and area-law entanglement accompanied by a qualitative change in long-distance mutual information near the SU(2)-symmetric point. 

Finally, several phases exhibit striking and potentially useful combinations of properties, as seen schematically in Fig.~\ref{subfig:phases-overview}. In particular, we identify in dense, long-range and single-basis circuits, rapidly purifying, non-scrambling regimes with long-range entanglement --including a phase that purifies on an \(O(1)\) timescale while remaining volume-law entangled, suggesting a measurement-only route to producing extensive long-range entanglement, e.g.,~for entanglement distribution between distant points. The absence of scrambling ensures that this entanglement remains locally accessible, and can be harvested without global operations. The \(O(1)\) purification timescale implies that the circuit reaches its entangled steady state in a depth independent of system size, so the measurement-only preparation protocol requires only \(O(N)\) total measurements, which is a favorable scaling for resource state generation. Moreover, fast purification ensures robustness to initial-state imperfections, since any residual correlations with an environment are rapidly erased by the measurement dynamics. Furthermore, we find that similar qualitative features (rapid purification, non-scrambling, and long-range correlations) also appear in shorter-range and sparser single-basis circuits, albeit with sub-volume entanglement scaling. While this phase offers a smaller entanglement resource, it requires only short-range parity checks, possibly making it more accessible to certain platforms.

This paper is organized as follows: In Sec.~\ref{Sec1}, we introduce the MoC models studied in this work, and their defining parameters as well as the trajectory-averaged observables we monitor in simulation together with our theoretical framework. In Sec.~\ref{Sec2}, we delve into the rich phases of prepared states produced by the MoCs as the circuit parameters are varied both in random- and single-basis designs. Finally, Sec.~\ref{Sec3} summarizes our results and discusses potential future directions.

\section{Models and preliminary concepts \label{Sec1}}
 In the following subsections, we introduce the family of circuits we study and provide a brief background on the theoretical analyses performed in this work.

\subsection{Circuit Setup}

\begin{figure}[!htb]
\centering
\subfloat[][]{
\includegraphics[width=1.0\columnwidth]{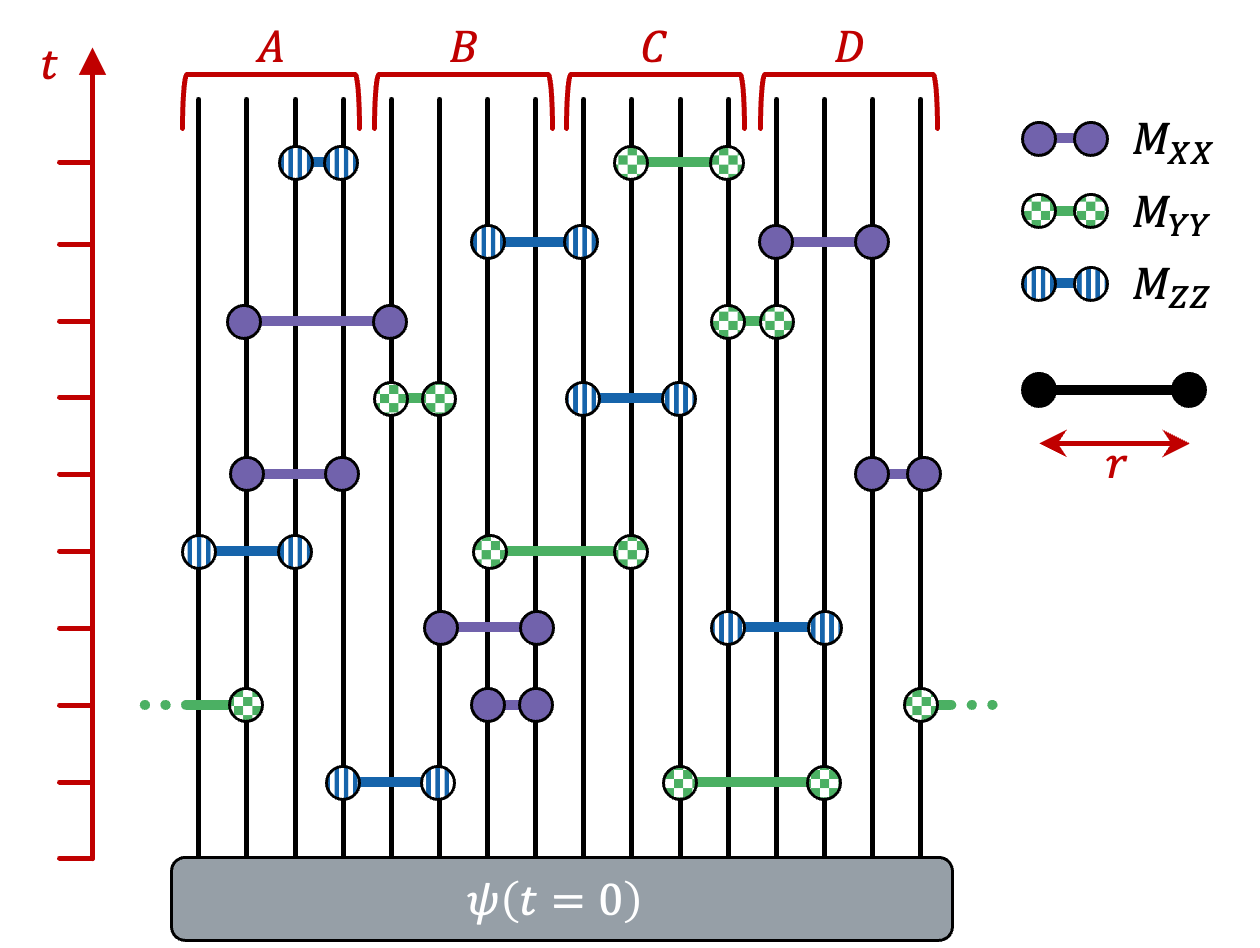}
\label{subfig:circ-setup}
}

\subfloat[][]{
\includegraphics[width=0.82\columnwidth]{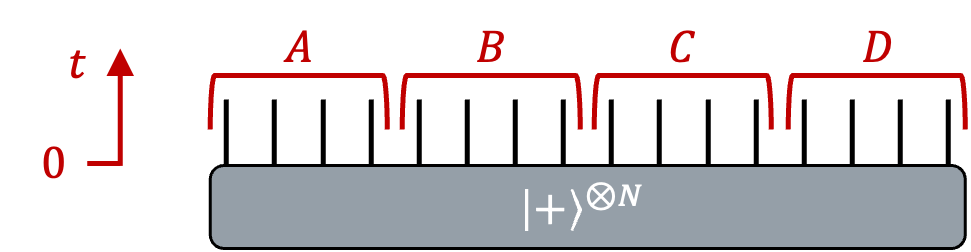}
\label{subfig:pure-init}
}

\subfloat[][]{
\includegraphics[width=0.87\columnwidth]{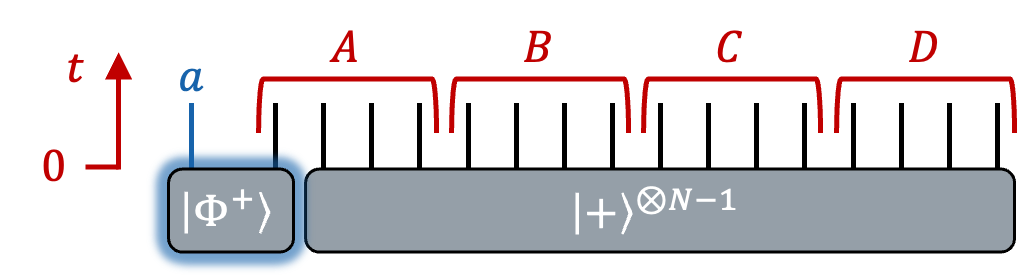}
\label{subfig:mixed-init}
}
\caption{\ref{subfig:circ-setup}
The long-range measurement-only circuit setup in our Clifford circuit simulations. The entire circuit is divided into four equal subsystems, named from $A$ to $D$. The circuit depth is depicted with a timeline marked by $t$ on the left. A circuit layer is a time step where $M_2$ many measurements are performed, and the circuit density in this schematic is $M_2/N=2/16=1/8$. In the random-basis measurement model, we sample the measurement basis randomly from the set $\{ X_{i}X_{j}, Y_{i}Y_{j}, Z_{i}Z_{j} \}$ for each measurement within a time step, whereas in the single-basis measurement model, the measurement basis is set to be one of $\{ X_{i}X_{j}, Y_{i}Y_{j}, Z_{i}Z_{j} \}$ for all measurements comprising one time step. The schematic highlights the random-measurement model. The distance $r$ of each measurement is determined by the $r^{-\alpha}$ probability distribution at a fixed interaction exponent $\alpha$. \ref{subfig:pure-init} Pure state initialization used for all simulations, except purification time analysis. Each qubit is initialized into the x-basis state $|+\rangle$. \ref{subfig:mixed-init} Mixed state initialization used for purification time analysis. An ancilla qubit $a$ and one qubit in the system are prepared as a Bell pair. }
\label{fig:rb-circ}
\end{figure}

We consider randomly generated MoCs comprised of two-qubit projective measurements. In particular, we study how the spatial distance between measurement sites, the measurement basis, and the density of these measurements per layer influence the properties of the states at the circuit output. The measurement operators are selected from the set $\{ X_{i}X_{j}, Y_{i}Y_{j}, Z_{i}Z_{j} \}$. Such a circuit involves only Clifford operations 
and can therefore be efficiently simulated with classical resources. 
Moreover, although these measurements are projective, they are informationally incomplete in the sense that the outcome of a two-qubit measurement does not fully determine the output state of the two involved qubits. Successive noncommuting measurements can therefore lead to larger entanglement within the system. 

Our system consists of $N$ qubits divided into four subsystems of equal size labeled $A$, $B$, $C$, and $D$ (see Fig.~\ref{subfig:circ-setup}). For the majority of simulations, we initialize the system in the $x$-basis product state $|\psi(t=0)\rangle = |+\rangle^{\otimes N}$ as illustrated in Fig. \ref{subfig:pure-init}. We then apply a randomly generated MoC whose particular structure is dictated by a few parameters: circuit density, measurement range, and layer-wise measurement basis. 

The circuit density $M_2 / N$ prescribes the number of two-qubit projective measurements, $M_2$, performed in each circuit layer. The term ``circuit layer'' here refers to each time step in which $M_2$ measurements are performed, see Fig.~\ref{subfig:circ-setup}. We note that this is distinct from the number of circuit layers that might comprise the circuit if it were to be compiled, which generally would be a smaller number as two (or more) sequentially simulated circuit layers may fully commute with one another, and therefore be combined into a single executable circuit layer. The measurements in each circuit layer are sampled independently from each other: each circuit layer is generated with no prior or future knowledge of the layer applied before or after. At its minimum, the circuit density can approach zero, which we take to be the limit where each time step contains only one two-qubit projective measurement regardless of the size of $N$. At its maximum, the circuit density can take the value $M_2/N=0.5$, in which case each qubit is involved in a two-qubit projective measurement. In other words, within a layer we enforce that each qubit participates in at most one measurement, so that all measurements in a layer act on disjoint pairs and commute.

We parameterize the range of two-qubit measurements comprising the circuit with an exponent $\alpha \geq 0$. If qubit $i$ is randomly selected to participate in a two-qubit measurement, the second qubit $j$ is selected probabilistically according to a distribution weighted by $r^{-\alpha}$, where $r$ is the distance between qubits $i$ and $j$ and periodic boundary conditions are assumed. A small value for $\alpha$ leads to long-range measurements, while a large value of $\alpha$ preferentially selects short-range measurements. The limit $\alpha = 0$ results in a uniform distribution where each qubit $j$ is just as likely to be selected regardless of range $r$. We note that due to the random circuit generation and the constraint that each qubit participate in at most one measurement within a given layer, a small number of long-range measurements may be included in a circuit layer even for large $\alpha$ if the circuit density is high. 

We examine two cases for layer-wise measurement basis. First, we consider the unconstrained case where each measurement operator is selected randomly and independently from the set $\{ X_{i}X_{j}, Y_{i}Y_{j}, Z_{i}Z_{j} \}$; this scenario is referred to as the \textit{random-basis case}. Secondly, we consider a constrained case where a single measurement operator is selected and consistently applied for each measurement comprising one circuit layer. We refer to this scenario, which is more experimentally feasible, as the \textit{single-basis case}. Fig. \ref{subfig:circ-setup} depicts an example circuit in which random-basis measurements are applied to 16 qubits. In this example, the circuit density $M_2 / N$ is set to $0.125$ and $\alpha$ has a large value ($\alpha \sim 3$), resulting in two short-range measurements in each circuit layer. The circuit layers are applied sequentially, and the dimension corresponding to circuit depth is denoted by $t$.  

As we explore the parameter space, we compute three quantities: the entanglement entropy $S$, the mutual information $\mathcal{I}$ between two distant qubits, and the tripartite mutual information $\mathcal{I}_3$ between three subsystems $A$, $B$, and $C$ as seen in Fig. \ref{subfig:circ-setup}. The entanglement entropy of a subsystem $a$ is computed as:
\begin{equation}
    S_{a} = - \text{Tr}\left( \rho_{a} \ln \rho_{a} \right).
\end{equation}
In circuit \ref{subfig:circ-setup}, we consider the union of subsystems $AB$ and we average over circuit trajectories $l$: $S = \langle S^{(l)}_{AB} \rangle_{l}$ capturing the entanglement of $AB$ with $CD$. The scaling of the entanglement entropy in system size and time informs us about the entropy production rate at the cut.  

The other quantities of interest can be computed from the entanglement entropy of different subsystems. The mutual information measures the amount of information that is shared between two subsystems, 
\begin{equation}
    \mathcal{I}(a;b) = S_{a} + S_{b} - S_{ab} .
\end{equation}
Mutual information is always nonnegative and entanglement between $a$ and $b$ will generate $\mathcal{I}(a;b) > 0$.
In circuit \ref{subfig:circ-setup}, we calculate the mutual information between two maximally distant qubits ($q_0$ and $q_{N/2-1}$) of each circuit realization initiated in the pure state $|+\rangle^{\otimes N}$. We average over circuit trajectories labeled by $l$ and define $\mathcal{I} = \langle \mathcal{I}^{(l)}(q_0; q_{N/2-1}) \rangle_l$. Therefore, a nonzero $\mathcal{I}$ along a given quantum trajectory indicates that chosen qubits share correlations, i.e.,~long-range and resourceful entanglement in the pure state wave function on that trajectory. In some cases, nonzero mutual information would indicate a finite probability of finding a Bell cluster residing at maximally distant qubits $q_0$ and $q_{N/2-1}$ \cite{PhysRevB.102.094204,PhysRevB.100.134306}. We calculate the number of Bell pairs across the system in addition to the mutual information. 

The tripartite mutual information across three regions, $a,b,c$, is given by:
\begin{equation}
    \mathcal{I}_3(a; b; c) = \mathcal{I}(a; b) + \mathcal{I}(a; c) - \mathcal{I}(a; bc)
\end{equation}
Tripartite mutual information quantifies how information from subregion $a$ is spread across regions $b$ and $c$ \cite{iyoda_scrambling_2018,PhysRevX.10.041020,kuno_phase_2023}. When information is delocalized across the system, the composite system $bc$ contains information about region $a$ that is otherwise inaccessible by regions $b$ and $c$ individually, leading to $\mathcal{I}(a; bc) > \mathcal{I}(a; b) + \mathcal{I}(a; c)$ and a negative value for $\mathcal{I}_3(a; b; c)$. Thus, the sign of $\mathcal{I}_3$ can be used to probe information scrambling \cite{iyoda_scrambling_2018}. We 
study the trajectory-averaged quantity $\mathcal{I}_3 = \langle \mathcal{I}^{(l)}_3(A; B; C) \rangle_l$ in circuit \ref{subfig:circ-setup}, where regions $A$, $B$, and $C$ each have size $N/4$.

Additionally, we quantify the purifying potential of the family of circuits when the system is initialized in a mixed state. One qubit within the system is entangled with an ancilla qubit outside the system at $t = 0$ to create a mixed state, as illustrated in Fig. \ref{subfig:mixed-init}. We then monitor the entanglement entropy between the ancilla qubit $q$ and the $N$ qubits comprising the system as the $N$ qubits are evolved under a random circuit and averaged over trajectories: $S_{q} = \langle S^{(l)}_{q} \rangle_l$. We assume an exponential decay in $S_{q} \sim e^{-t/\tau}$ \cite{PhysRevX.10.041020,majidy_critical_2023} and fit the resulting curves to determine the purification timescale $\tau$. In particular, we are interested in whether a circuit can purify the state within a purification timescale $\tau(N)$ that grows less than exponentially in the system size. 

\subsection{\label{sec:methods}Methods for mapping MoC to statistical mechanics model and an effective Hamiltonian}

In this section we review the mapping methods utilized in the paper. We start with the mapping of the entanglement entropy in the 1+1D random-basis MoC setup to free energy in a 2D statistical mechanics model under certain assumptions to arrive at insights for the numerical results presented in the next section. Such mappings have been shown to be useful for random unitary circuits \cite{PhysRevB.101.104301, PhysRevB.101.104302, PhysRevLett.128.010604}. Then, we take the continuous time limit of the statistical mechanics model that we derived and obtain an effective Hamiltonian \cite{PhysRevLett.128.010604}, which is found to be a long range interacting XX chain model.

\subsubsection{Trajectory Averaged Entanglement Entropy}\label{entropy_derivation}

Here we express $\langle S^{(l)}_{A} \rangle_l$, the von Neumann entanglement entropy of subsystem $A$ averaged over all trajectories and circuit realizations.
Let $C$ denote the Kraus operator associated with a given circuit instance and a particular set of measurement outcomes. We define $\mathbb{E}_C$ as an expectation value over all Kraus operators $C$, with measurement locations and bases sampled uniformly and measurement outcomes weighted by their Born probabilities (which, for our circuits, are all equal to $1/2$). The average von Neumann entropy of the final state after evolving under random MoCs can be found by evaluating the entropy for the properly normalized evolved state $C|\psi\rangle/\|C|\psi\rangle\|$ and averaging over trajectories \cite{PhysRevB.101.104302,Stabilizer_Entropy,PhysRevB.94.014205,PhysRevB.103.L100207}:
\begin{equation}\label{eq:entropy-def}
    \langle S^{(l)}_{A} \rangle_l
    =\underset{C}{\mathbb{E}} \left[ S_{ A}\left(\frac{C|\psi\rangle}{\| C|\psi\rangle \|}\right) \times \operatorname{Tr}\left(C|\psi\rangle\langle\psi| C^{\dagger}\right)\right], 
\end{equation}
where $\operatorname{Tr}\left(C|\psi\rangle\langle\psi| C^{\dagger}\right)$ is the Born probability of each trajectory $C$. Eq.~\eqref{eq:entropy-def} can be recast using the so-called replica trick which was introduced for random tensor networks and unitary circuits by \cite{PhysRevB.99.174205,PhysRevB.100.134203}. A replica index $n$ is introduced to treat the logarithm in the formula for the von Neumann entropy $S_A$.
See the full derivation in ~\cite{PhysRevB.101.104301} to obtain $\langle S^{(l)}_{A} \rangle_l$ as a limit of $n^{\rm th}$ order ``conditional Rényi entropies" that is denoted below by $\tilde{S}^{(n)}_A$:
\begin{align}\label{eq:conditional_renyi}\langle S^{(l)}_{A} \rangle_l = \lim_{n \rightarrow 1} \tilde{S}^{(n)}_A \equiv \lim_{n \rightarrow 1} \frac{\log(\mathcal{Z_A}^{(n)})-\log(\mathcal{Z_{\emptyset}}^{(n)})}{1-n},\end{align}
where the quantities $\mathcal{Z_A}$ and $\mathcal{Z_{\emptyset}}$ can be expressed as:
\begin{align}\label{eq:intermsofswqp}
    \mathcal{Z_A}^{(n)} &= \underset{C}{\mathbb{E}} \Tr(C\ketbra{\psi}{\psi}C^{\dag})^{\otimes n} \mathcal{S}_{n,A}) \\
    \mathcal{Z_{\emptyset}}^{(n)} & = \underset{C}{\mathbb{E}} \Tr(C\ketbra{\psi}{\psi}C^{\dag}))^{\otimes n}. \notag
\end{align}
Here, $\mathcal{S}_{n,A}$
is the permutation operator that acts on $n$ replicas of the system and is defined to permute only the replicas of the subsystem $A$. $\mathcal{S}_{n, A}$ factorizes over sites as:
\begin{equation*}
\mathcal{S}_{n, A}=\prod_x   \chi_{g_x}, 
\end{equation*}
where $\chi_{g_x}$ is the operator that permutes the $n$ replicas of the qubit $x$ according to the permutation rule $g_x$:
\begin{eqnarray}
    \chi_{g_x} &=& \sum_{i_1,\ldots,i_n}\left|i_{g_x(1)} i_{g_x(2)} \cdots i_{g_x(n)}\right\rangle\left\langle i_1 i_2 \cdots i_n\right|, \label{eq:permOperators}\\ 
    g_x &=& \begin{cases} (12 \cdots n), & x \in A \\ \text { identity }=e, & x \in \bar{A}.\notag\end{cases}
\end{eqnarray}
Since the local Hilbert space is a qubit, $i_k \in \lbrace 0, 1 \rbrace$ where the subscript $k$ is the replica label. The notation $(12 \cdots n)$ is the $n$-cycle; equivalently, $\chi_{g_x}$ relabels replicas by $k\mapsto g_x(k)$ on site $x$. For $n=2$, $g_x=(12)$ for \(x\in A\) and \(g_x=e\) for \(x\in\bar A\), so sites in \(A\) swap the two replicas while sites in \(\bar A\) act trivially.

In the next subsection, we will see that the quantities $\mathcal{Z_A}$ and $\mathcal{Z}_{\emptyset}$ can be associated with partition functions of a 2D model. This result suggests that computing the Rényi entropy is equivalent to finding the free energy cost of changing a partition function from $\mathcal{Z}_{\emptyset}$ to $\mathcal{Z}_A$. With this interpretation in mind, in what follows we extend the methodology to handle the two-qubit measurements that appear in our circuit setup, ultimately mapping our family of MoCs to a statistical mechanics model.

\subsubsection{Random-basis statistical mechanics model \label{sec:randomBasisStatMech}}

Here we formulate the functions $\mathcal{Z}_A$ and $\mathcal{Z}_{\emptyset}$ in the random-basis MoC as partition functions of a classical statistical mechanics model and the entropy as the free energy cost associated with changing a boundary condition. Similar mappings have been done for random unitary circuits with single qubit measurements \cite{PhysRevB.101.104301, PhysRevB.101.104302, PhysRevLett.128.010604}, but here we generalize the idea to the case of random-basis MoC with 2-qubit measurements. 

Let us consider as an example a random-basis  circuit with $M_2/N = 0.5$, the leftmost diagram below:
\begin{equation}
\includegraphics[width=\columnwidth]{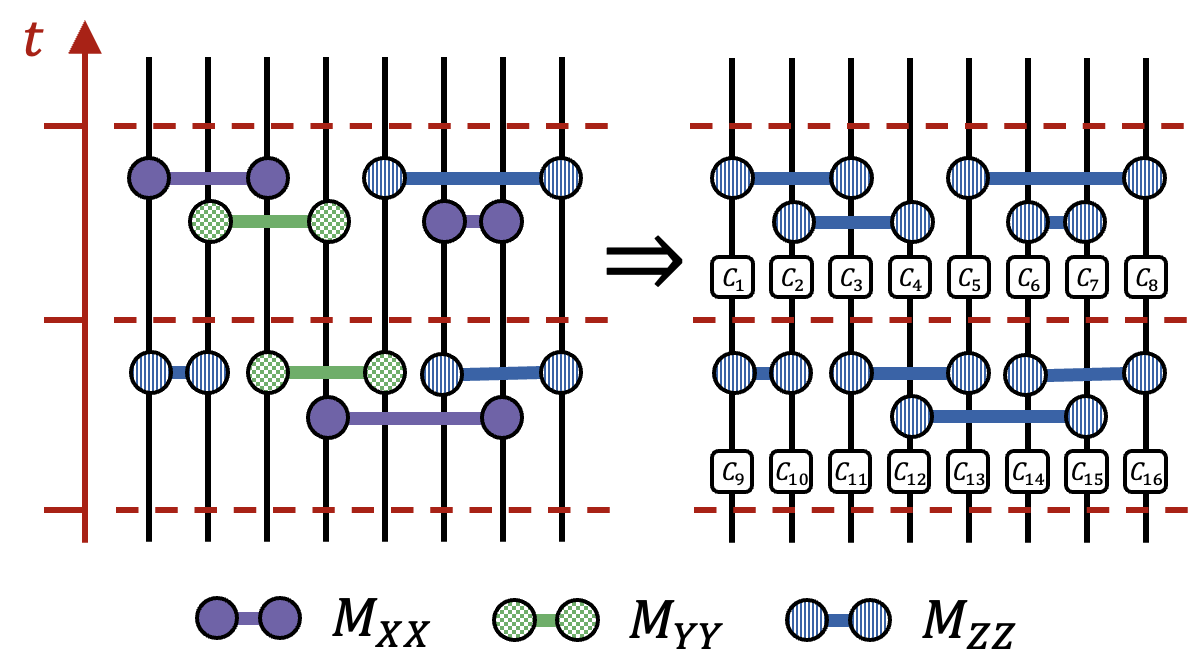}
\label{subfig:mapping-a}
\end{equation}
Conjugating a Z measurement by a 
single-qubit Clifford gives either an X, Y or Z measurement with equal probability. Thus, the different basis measurements that appear in the circuit can be transformed to be $M_{\rm ZZ}$ by introducing single qubit Clifford gates between layers in the circuit, hence changing the basis before each measurement layer (rightmost diagram in Eq.~\eqref{subfig:mapping-a}). It is important to note that due to the structure of the circuit, i.e.,~having either of $XX, YY$ or $ZZ$ measurements, despite a random choice of basis, inserted single-qubit Clifford gates will not be fully independent within a layer. However, in the analytical calculation below we assume that all single-qubit Clifford gates are independent and uniformly sampled from the Clifford group. This assumption allows for replacing Clifford gates with Haar random unitary gates in the calculation of $\mathcal{Z}_A$ and $\mathcal{Z}_{\emptyset}$, given the fact that Clifford circuits are 3-design for replica number $n \le 3$, which includes the replica limit $n \rightarrow 1$ \cite{Webb2015}. The reasons for such an assumption are (i) circuits with the Haar ensemble are analytically tractable as shown below, and (ii) the discrete basis can be considered as a microscopic limit of the continuous Haar basis revealing certain universal properties of the structured circuit.

Taking the expectation value over $C$ in the partition functions (Eqs.~\eqref{eq:intermsofswqp}) leads to taking the ensemble average over all single qubit Haar random unitaries. Haar averaging of the partition functions for a single layer circuit with $n=3$ replicas can be diagrammatically depicted as:
\begin{equation}
\includegraphics[width=1\columnwidth]{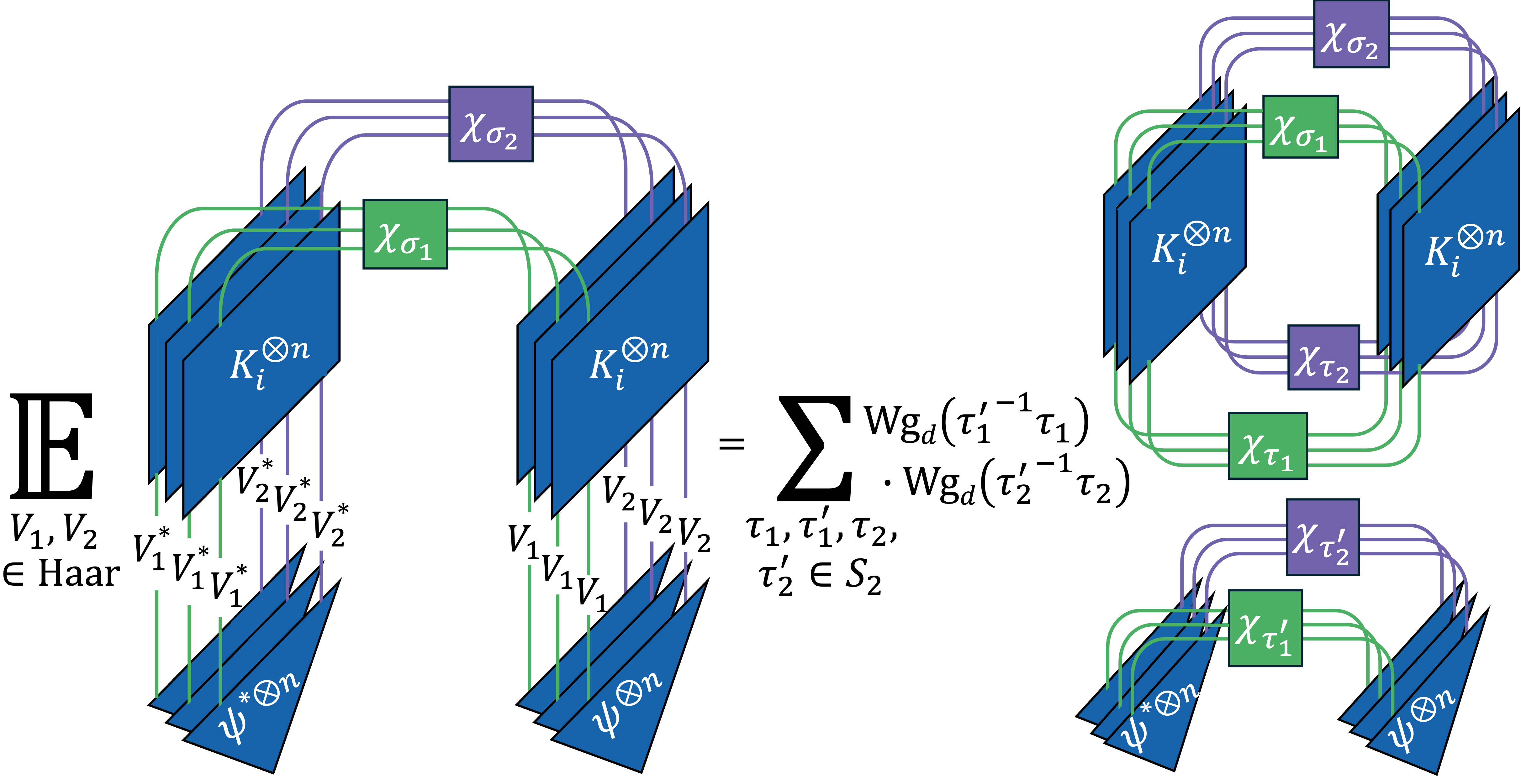}\label{subfig:rb-haar}
\end{equation}
where $\mathrm{Wg}_d(\tau)$ is the Weingarten function of permutation $\tau$ in $d-$dimension, $\chi_{\tau_1}, \chi_{\tau_2}, \chi_{\tau_1^{'}} \dots$ etc. are permutation operators on the replicas as in Eq.~\eqref{eq:permOperators}, and $K_i^{\otimes n}$ denote the Kraus operators $K_i$ in $n-$copy replicated space corresponding to $M_{\rm ZZ}$ in Eq.~\eqref{subfig:mapping-a}. A natural generalization of the \(ZZ\) parity measurement is the mod$-d$ parity measurement with \(d\) Kraus operators
\begin{equation}
K_i := \sum_{j=0}^{d-1}\ket{j,\;j+i}\bra{j,\;j+i},
\quad i\in\{0,1,\dots,d-1\},
\end{equation}
where the addition \(j+i\) is understood modulo \(d\).
For qubits (\(d=2\)), these are precisely the even/odd parity projectors:
\(K_0=\ket{00}\bra{00}+\ket{11}\bra{11}=(I+ZZ)/2\) and
\(K_1=\ket{01}\bra{01}+\ket{10}\bra{10}=(I-ZZ)/2\). Using Eq.~\eqref{subfig:rb-haar}, the Haar averaging for a multilayer circuit can be done iteratively one layer at a time.

This procedure gives rise to a classical statistical model in two dimensions for $\mathcal{Z}_A$ and $\mathcal{Z}_{\emptyset}$ whose lattice geometry is visible in the right hand side of the following equation.
\begin{equation}
\includegraphics[width=0.7\columnwidth]{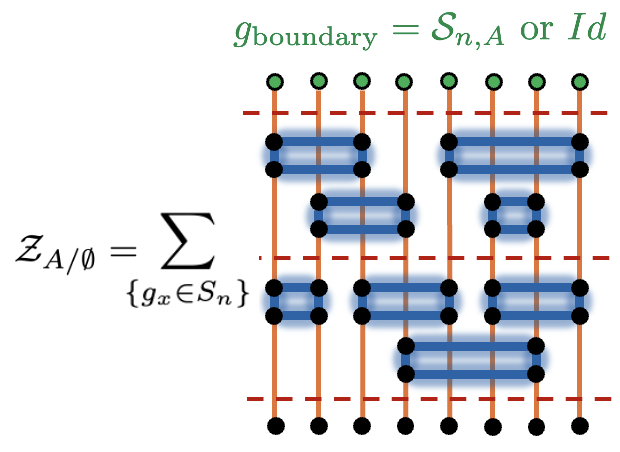} 
\label{subfig:mapping-c}
\end{equation}
Each black dot in Eq.~\eqref{subfig:mapping-c} carries a permutation variable \(g_x\in S_n\) (equivalently a ``spin'' taking values in \(S_n\) \cite{PhysRevB.101.104302}), corresponding to the replica permutation operator \(\chi_{g_x}\) in the replicated Hilbert space. The partition function is obtained by summing over all assignments \(\{g_x\}\). The orange links denote Weingarten weights \(\mathrm{Wg}_d(g_x^{-1}g_{x'})\) that arise from averaging the single-site unitary gates in Eq.~\eqref{subfig:rb-haar} and couple neighboring permutation variables. Each blue highlighted rectangle denotes an interaction term $W_M$, i.e.,~the local measurement weight, arising due to two-qubit measurements, written diagrammatically for an $n=3$-replica circuit as:
\begin{equation}
\includegraphics[width=0.8\columnwidth]{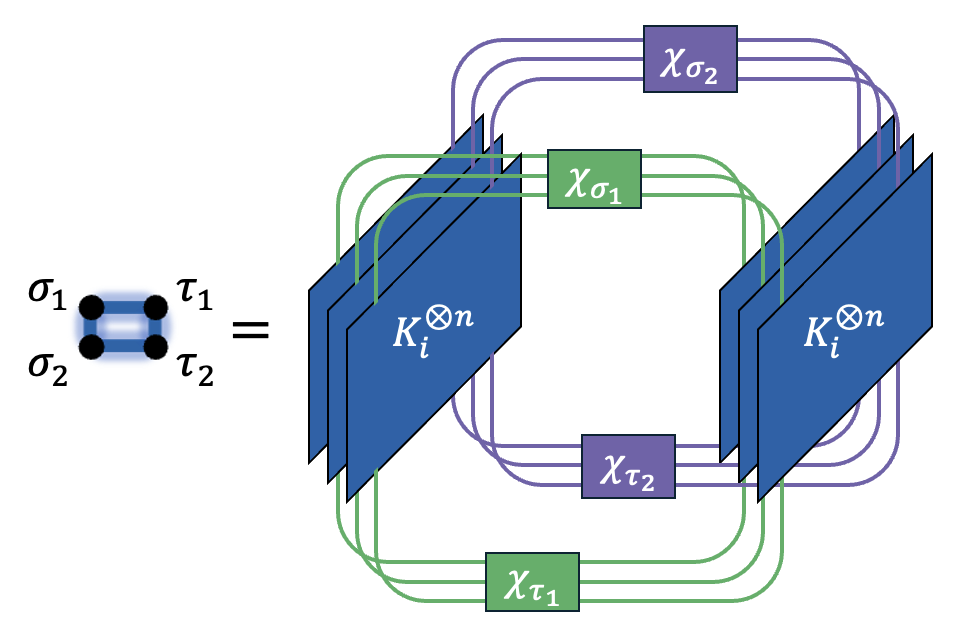}
\label{subfig:mapping-d}
\end{equation}
Algebraically, Eq.~\eqref{subfig:mapping-d} reads,
\begin{align}
    W_M &\left( \sigma_1, \sigma_2, \tau_1, \tau_2 \right)\label{eq:localMeasurementWeight}\\ &= \sum_{i = 0}^{d-1} \operatorname{Tr} \left( K_{i}^{\otimes n} (\chi_{\sigma_1} \otimes \chi_{\sigma_2}) K_i^{\otimes n} (\chi_{\tau_1} \otimes \chi_{\tau_2})\right)\notag.
\end{align}
Therefore, Eq.~\eqref{subfig:mapping-c} specifies the lattice geometry of the resulting two-dimensional statistical mechanics model, with black dots, orange links and blue rectangles representing lattice sites, couplings and interactions, respectively.

The partition function is then the sum of Eq.~\eqref{subfig:mapping-d} over all possible permutations defined at each site leading to Eq.~\eqref{subfig:mapping-c}. This algebraically is:
\begin{eqnarray}\label{eq:rand_basis_partition_funct}
    \mathcal{Z}_{\emptyset/\mathcal{A}} &=& \sum_{\left\{\sigma_i \in S_Q\right\}} \prod_{\langle i j k \ell \rangle \in \mathcal{E}_4} W_M\left(\sigma_i, \sigma_j, \sigma_k, \sigma_{\ell} \right) \notag \\
    &\times &\prod_{\langle i j\rangle \in \mathcal{E}_2} \mathrm{Wg}_{d}\left(\sigma_i^{-1} \sigma_j\right),
\end{eqnarray}
where $\mathcal{E}_4$ denotes the interaction terms, i.e.,~blue rectangles, and $\mathcal{E}_2$ denotes the coupling terms, i.e.,~orange links, as shown in Eq.~\eqref{subfig:mapping-c}.

The boundary condition at the final time edge, represented by the green dots in Eq.~\eqref{subfig:mapping-c}, are fixed by the expressions for $\mathcal{Z}_{\emptyset}$ and $\mathcal{Z_{A}}$ in Eq \eqref{eq:intermsofswqp}. The boundary permutations in the calculation of $\mathcal{Z}_{\emptyset}$ are all pinned to be equal to the identity, while the insertion of \(\mathcal{S}_{n,A}\) in $\mathcal{Z}_{A}$ switches the boundary permutation from the identity to the replica swap permutation on region \(A\) such that the entanglement entropy is the free-energy cost associated with this change in boundary condition.

Next we evaluate the local measurement weights $W_M(\sigma_1,\sigma_2,\tau_1,\tau_2)$ associated with a two-site projective measurement block in the circuit for general qudit dimension $d$ and replica number $n$. We find that $W_M$ is larger when the four permutations ``agree'',  as explained in the following section. 

To evaluate \(W_M\), we expand each \(K_i^{\otimes n}\) in Eq. \eqref{eq:localMeasurementWeight} in the product basis.
We introduce replica indexed labels \(j_1,\dots,j_n\) and \(k_1,\dots,k_n\) in \(\{0,\dots,d-1\}\), so that:
\begin{widetext}
\begin{align}
W_M(\sigma_1,\sigma_2,\tau_1,\tau_2)
&=\sum_{i=0}^{d-1}\;
\sum_{\substack{j_1,\ldots,j_n\\ k_1,\ldots,k_n}}
\Bigl\langle j_1,j_1+i;\,\ldots;\,j_n,j_n+i \Big|
(\chi_{\sigma_1}\otimes \chi_{\sigma_2})
\Big| k_1,k_1+i;\,\ldots;\,k_n,k_n+i \Bigr\rangle \nonumber\\
&\hspace{2.5cm}\times
\Bigl\langle k_1,k_1+i;\,\ldots;\,k_n,k_n+i \Big|
(\chi_{\tau_1}\otimes \chi_{\tau_2})
\Big| j_1,j_1+i;\,\ldots;\,j_n,j_n+i \Bigr\rangle .
\label{eq:WM_expanded_main}
\end{align}
\end{widetext}
Using our convention for \(\chi_g\), the matrix elements enforce Kronecker-delta constraints that identify replica indices:
\begin{align}
& (\chi_{\sigma_1}\otimes \chi_{\sigma_2})
\ket{k_1,k_1+i;\,\ldots;\,k_n,k_n+i} \notag \\
&=
\ket{k_{\sigma_1(1)},k_{\sigma_2(1)}+i;\,\ldots;\,k_{\sigma_1(n)},k_{\sigma_2(n)}+i},\label{eq:perm_action_main}
\end{align}
and similarly for \(\tau_1,\tau_2\).
Therefore a term in Eq.~\eqref{eq:WM_expanded_main} is nonzero only if, for all replica labels \(m\in[n]\),
\begin{align}\label{eq:constraints_main}
j_m &= k_{\sigma_1(m)},\quad
j_m = k_{\sigma_2(m)},\\
k_m &= j_{\tau_1(m)},\quad
k_m = j_{\tau_2(m)} , \notag
\end{align}
where the second and fourth equalities follow because the ``\(+i\)'' shifts cancel modulo \(d\).
Eliminating the \(k\)'s from Eq.~\eqref{eq:constraints_main} yields constraints solely among the \(j\)'s:
\begin{equation}\label{eq:j_constraints_main}
j_m = j_{(\tau_1\sigma_1)(m)},\quad
j_m = j_{(\tau_2\sigma_1)(m)},\quad
j_m = j_{(\tau_2\sigma_2)(m)}, \notag
\end{equation}
for all \(m\in[n]\).
Equivalently, \(j_m\) must be constant on orbits of the subgroup
\begin{equation}
G:=\langle \tau_1\sigma_1,\;\tau_2\sigma_1,\;\tau_2\sigma_2\rangle \le S_n .
\end{equation}
Since each orbit can be assigned an independent value in \(\{0,\dots,d-1\}\), the number of admissible assignments of \(\{j_m\}\) is \(d^{\#\mathrm{Orb}(G)}\).
Finally, the Kraus label \(i\in\{0,\dots,d-1\}\) is unconstrained and contributes an additional factor of \(d\).
We thus obtain the compact orbit-counting form:
\begin{equation}\label{eq:WM_orbit_main}
W_M(\sigma_1,\sigma_2,\tau_1,\tau_2)
= d^{\,1+\#\mathrm{Orb}\!\left(\langle \tau_1\sigma_1,\;\tau_2\sigma_1,\;\tau_2\sigma_2\rangle\right)} .
\end{equation}
Using elementary generator manipulations, this may also be written as:
\begin{equation}
W_M(\sigma_1,\sigma_2,\tau_1,\tau_2)
= d^{\,1+\#\mathrm{Orb}\!\left(\langle \tau_1\tau_2^{-1},\;\tau_2\sigma_1,\;\sigma_1^{-1}\sigma_2\rangle\right)} , \notag
\end{equation}
which is sometimes more convenient in practice.
Eq.~\eqref{eq:WM_orbit_main} makes explicit that \(W_M\) is largest when the permutations ``agree'' (yielding more orbits), and smaller when the permutations strongly mix replica labels (fewer orbits), thereby penalizing domain-wall crossings in the associated statistical mechanics model.

When $n=2$ the formula is particularly simple: it says $W_M(\sigma_1,\sigma_2,\tau_1,\tau_2) = d^3$ for $\sigma_1=\sigma_2=\tau_1=\tau_2$ and $d^2$ otherwise. This means that each block $W_M$ crossing the identity and swap domains in the calculation of entanglement entropy adds some free energy cost to the partition function, thus increasing the entanglement entropy.

\subsubsection{Mapping to an effective XX hamiltonian \label{ref:effectiveXYphase}}

Adapting the methodology of \cite{PhysRevLett.128.010604}, which writes an effective Hamiltonian for long-range unitary circuits with single qubit measurements, we show that the continuous time limit of the partition function in the random basis MoC, Eq.~\eqref{eq:rand_basis_partition_funct}, for $n=2$ replicas is equivalent to imaginary time evolution under an effective Hamiltonian. 

We start by writing $\tilde{S}^{(n)}_A$, the $n^{\rm th}$ order conditional Renyi entropy (Eq.~\eqref{eq:conditional_renyi}), in the the duplicated Hilbert space notation as:
\begin{align*}
    \tilde{S}^{(n)}_A = \frac{1}{1-n} \log \left(\frac{\left.\left\langle\langle\mathcal{I}| \mathcal{S}_{n, A} | \:\rho^{(n)}\right\rangle\right\rangle}{\left\langle\left\langle\mathcal{I} |  \rho^{(n)}\right\rangle\right\rangle}\right),
\end{align*}
where the duplicated Hilbert space is $\mathcal{H}^{(n)}:= (\mathcal{H}\otimes \mathcal{H^*})^{\otimes n}$, the vector $|\rho^{(n)}\rangle\rangle \in \mathcal{H}^{(n)}$ and $|\mathcal{I}\rangle \rangle$ represent $(C\ketbra{\psi}{\psi}C^{\dag})^{\otimes n}$ and the identity operator, respectively, so that $\langle \langle \mathcal{I}|\rho^{(n)}\rangle\rangle = \underset{C}{\mathbb{E}} \Tr (C\ketbra{\psi}{\psi}C^{\dag})^{\otimes n} = \mathcal{Z}_{\emptyset}$. $\mathcal{S}_{n, A}$, as defined before, permutes only the replicas of the subsystem $A$ acting on $n$ replicas. 

The random basis MoC can be expressed as an effective time evolution on $|\rho^{(n)}\rangle\rangle$. In the previous subsection, we had already rewritten the basis randomness as single-qubit random Cliffords preceding each measurement and approximated the Clifford average by a Haar average to access the projector structure. The resulting circuit consisted of alternating rows of single-qubit Haar unitaries and $Z_iZ_j$-basis measurements. For \(n=2\), the on-site Haar average over single qubit unitaries acts in the duplicated Hilbert space and projects onto a two-dimensional local subspace spanned by the identity and swap permutation sectors \(\{|\mathcal I\rangle\rangle,|\mathcal S\rangle\rangle\}\) \cite{PhysRevLett.128.010604}. Equivalently, the averaged on-site channel is exactly the projector
\begin{equation}
\mathcal U_{\mathrm{Haar}} \equiv \mathcal P
=\bigotimes_{j=1}^{L}\sum_{\sigma_j=\uparrow,\downarrow}
|\sigma_j\rangle\rangle\langle\langle \sigma_j|_j,
\label{eq:HaarProjector}
\end{equation}
where \(\{|\uparrow\rangle\rangle_j,|\downarrow\rangle\rangle_j\}\) is any orthonormal basis of the \(\{|\mathcal I\rangle\rangle_j,|\mathcal S\rangle\rangle_j\}\) subspace. We therefore identify this subspace with an effective spin-\(\tfrac12\) at each site and introduce Pauli operators \(\sigma_j^{x,z}\) acting on it (following Ref.~\cite{PhysRevLett.128.010604}).

Now let $\mathcal{M}_{ij}$ be the operator that acts in the duplicated Hilbert space as a $Z_iZ_j$ measurement on qubits $i$ and $j$. 
In the duplicated Hilbert space, the circuit is composed of adding and concatenating layers of the form
$\mathcal{P}\mathcal{M}_{ij}\mathcal{P},$ since $\mathcal{P}$ is simply a projection operator onto the subspace spanned by the permutations \(\{|\mathcal I\rangle\rangle_j,|\mathcal S\rangle\rangle_j\}\).
Notice that the projection of the $Z_iZ_j$ measurement onto this subspace has matrix elements given by the $W_M$ operator, that is:
\begin{align*}
    \langle \langle \sigma_i, \sigma_j| \mathcal{M}_{ij} | \tau_i, \tau_j \rangle \rangle = W_M(\sigma_i,\sigma_j, \tau_i, \tau_j),
\end{align*}
as can be seen from Eq.~\eqref{subfig:mapping-d}. For the case $n=d=2$ studied here, Eq.~\eqref{eq:WM_orbit_main} reduces to a particularly simple form: $\langle \langle \sigma_i, \sigma_j| \mathcal{M}_{ij} | \tau_i, \tau_j \rangle \rangle = 8$ for $\sigma_i=\sigma_j=\tau_i=\tau_j$ and $4$ otherwise.  Doing a change of basis to rewrite $\mathcal{M}_{ij}$ in terms of the effective Pauli basis $|+\rangle\rangle$ and $|-\rangle\rangle$, we obtain
\begin{align*}
    \mathcal{P}\mathcal{M}_{ij}\mathcal{P} = C + J(\sigma_i^z\sigma_j^z + \sigma_i^x\sigma_j^x),
\end{align*}
where $J$ and $C$ are normalization constants.

The time evolution of the density matrix under the random-basis long-range interacting MoC is then given by:
\begin{align*}
    \mathcal{P}|\rho&(t+\delta t) \rangle \rangle \\
    &= \left(C + J \sum_{i,j}|i-j|^{-\alpha} (\sigma_i^z\sigma_j^z + \sigma_i^x \sigma_j^x)\right)\mathcal{P}|\rho(t)\rangle\rangle.
\end{align*}
We take a continuous time limit in which the transfer matrix for a single circuit layer differs from the identity by \(O(\delta t)\), i.e.,
\begin{align*}
    \mathcal{P}|\rho(t+\delta t) \rangle \rangle = \bigg( \mathbf{1}-\delta t\,H_{\mathrm{eff}}+O(\delta t^{2})\bigg)\mathcal{P}|\rho(t)\rangle\rangle
\end{align*}
so that the coefficients appearing in the projected layer satisfy \(C=1+O(\delta t)\) and \(J=O(\delta t)\). Then the effective Hamiltonian reads,
$$H_{\rm eff} = - J \sum_{i,j}|i-j|^{-\alpha} (\sigma_i^z\sigma_j^z + \sigma_i^x \sigma_j^x),$$
up to an additive energy shift. 

This is the Hamiltonian for a long-range interacting XX chain in the $xz$-plane. Thus, the continuous limit of our circuit is expected to evolve to the ground state of this Hamiltonian, allowing us to compare observed properties of our circuit to those of the XX chain in the next section.

\section{Results and discussion}\label{Sec2}

\subsection{Random-basis measurement model} 

\begin{figure}[htb]
\centering
\includegraphics[width=1.0\columnwidth]{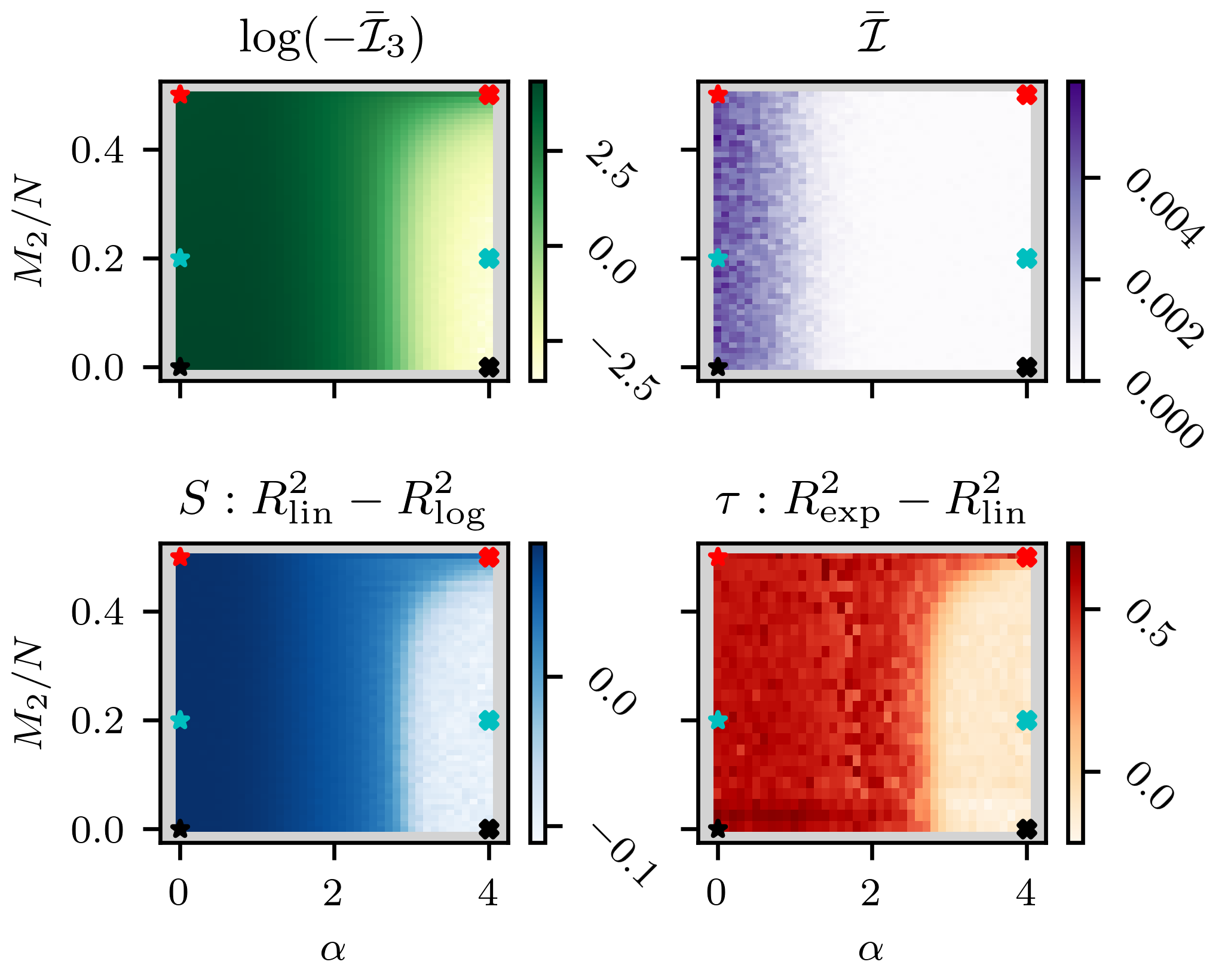}
\caption{ Phase diagrams for random-basis measurement scheme. Probabilistic measurement range decreases from left to right with increasing $\alpha$, and circuit density $M_2/N$ increases from bottom to top. Top left plot: logarithm of negated tripartite mutual information in the steady state. Top right plot: mutual information between two maximally distant qubits in the steady state. Bottom left plot: difference between fitting $R^2$ values when steady state bipartite entropy data vs system size $N$ is fit to a linear function and a logarithmic function. Positive regions (dark blue) indicate volume law entanglement ($R^2_{\text{lin}} > R^2_{\text{log}}$) while negative regions (light blue) indicate sub-volume law entanglement ($R^2_{\text{lin}} < R^2_{\text{log}}$). Bottom right plot: difference between fitting $R^2$ values when purification timescale $\tau$ vs system size $N$ is fit to an exponential function and a linear function. Positive regions (dark red) indicate non-purifying ($R^2_{\text{exp}} > R^2_{\text{lin}}$) while negative regions (yellow) indicate purifying ($R^2_{\text{exp}} < R^2_{\text{lin}}$). The $M_2 / N = 0$ limit (one measurement per layer regardless of $N$) is treated slightly differently: to account for the expected additional factor of $N$ in time, the values of $\tau$ are divided by $N$ prior to fitting.}
\label{fig:rb-pd}
\end{figure}

First, we examine the more general case of employing a random measurement basis chosen from $\{X_iX_j,Y_iY_j,Z_iZ_j\}$ for each two-qubit measurement. We vary the circuit density $0 \leq M_2 / N \leq 0.5$ and $0 \leq \alpha \leq 4$, considering systems that range in size from $N = 64$ to $N = 512$. At each point in the parameter space, we evolve 1000 randomly generated circuits for $2N^2 / M_2$ layers for each system size $N$ to ensure we reach steady state. At 100 intermediate points $t$ during the evolution, we compute $S$, $\mathcal{I}$, and $\mathcal{I}_3$. Fig. \ref{fig:rb-pd} displays the resulting phase diagrams across the parameter space. Note that we use a bar to indicate steady state values, which we define as the quantity averaged over both the 1000 circuit instances and the final 20 intermediate points $t$ to reduce sampling uncertainty. The random-basis phase diagrams are discussed in the following sections, and a summary of the results is provided in Table \ref{tab:rb}. 

\begin{table}[!htb]
    \begin{center}
    \caption{Summary of findings for random-basis MoCs where $\mathcal{I}_3$ is the tripartite mutual information, $\mathcal{I}$  is the mutual information, $S$ is the bi-partite entanglement entropy and $\tau$ is the purification time.}
    \label{tab:rb}
    %\begin{tabular}{|>{\bfseries}c|>{\centering\arraybackslash}m{1.2cm} | >{\centering\arraybackslash}m{1.2cm}|>{\centering\arraybackslash}m{1.3cm}|>{\centering\arraybackslash}m{1.3cm}|}
    \begin{tabular}{|c|c|c|c|c|}
        \hline \rule{0pt}{12pt}
        & \textbf{$\mathcal{I}_3$} & \textbf{$\mathcal{I}$} & \textbf{$S$} & \textbf{$\tau$} \\[2pt]
        \hline
        \makecell{\textbf{short range} \\  \textbf{sparse layers} \\ $\alpha \sim 4, M_2/N \sim 0$} & $\lesssim 0$ & $=0$ & $\mathcal{O}(\log N)$ & $\mathcal{O}(N)$ \\
        \hline 
        \makecell{\textbf{short range} \\ \textbf{dense layers} \\ $\alpha \sim 4, M_2/N \lesssim 0.5$} & $< 0$ & $=0$ & $\mathcal{O}(N)$ & $\mathcal{O}(e^N)$ \\
        \hline
        \makecell{\textbf{long range} \\ \textbf{sparse layers} \\ $\alpha \sim 0, M_2/N \sim 0$} & $<0$ & $\neq 0$ & $\mathcal{O}(N)$ & $\mathcal{O}(e^N)$ \\
        \hline
        \makecell{\textbf{long range} \\ \textbf{dense layer} \\ $\alpha \sim 0, M_2/N \lesssim 0.5$} & $<0$ & $\neq 0$ & $\mathcal{O}(N)$ & $\mathcal{O}(e^N)$ \\
        \hline
        \bottomrule
    \end{tabular}
    \end{center}
\end{table}

\subsubsection{Phases of Entanglement Growth}\label{entanglement_growth1}
In the lower left plot of Fig. \ref{fig:rb-pd}, we examine the entanglement entropy scaling with system size by comparing $R^2$ values when the $\bar{S}$ versus system size $N$ data is fit to a linear and a logarithmic function. Linear entanglement growth is indicative of a volume-law entangled phase, while logarithmic entanglement growth (also referred to as sub-volume law) is typically associated with the critical transition between volume and area-law entanglement in a measurement-induced phase transition (MIPT) \cite{PhysRevX.9.031009, PhysRevB.101.104302, lavasani_measurementinduced_2021}. For the family of measurement-only circuits considered here, we observe no area-law entanglement; only linear and logarithmic entanglement growth is observed. 

\begin{figure*}[!htb]
\centering
\subfloat[][]{
\includegraphics[width=1.0\columnwidth]{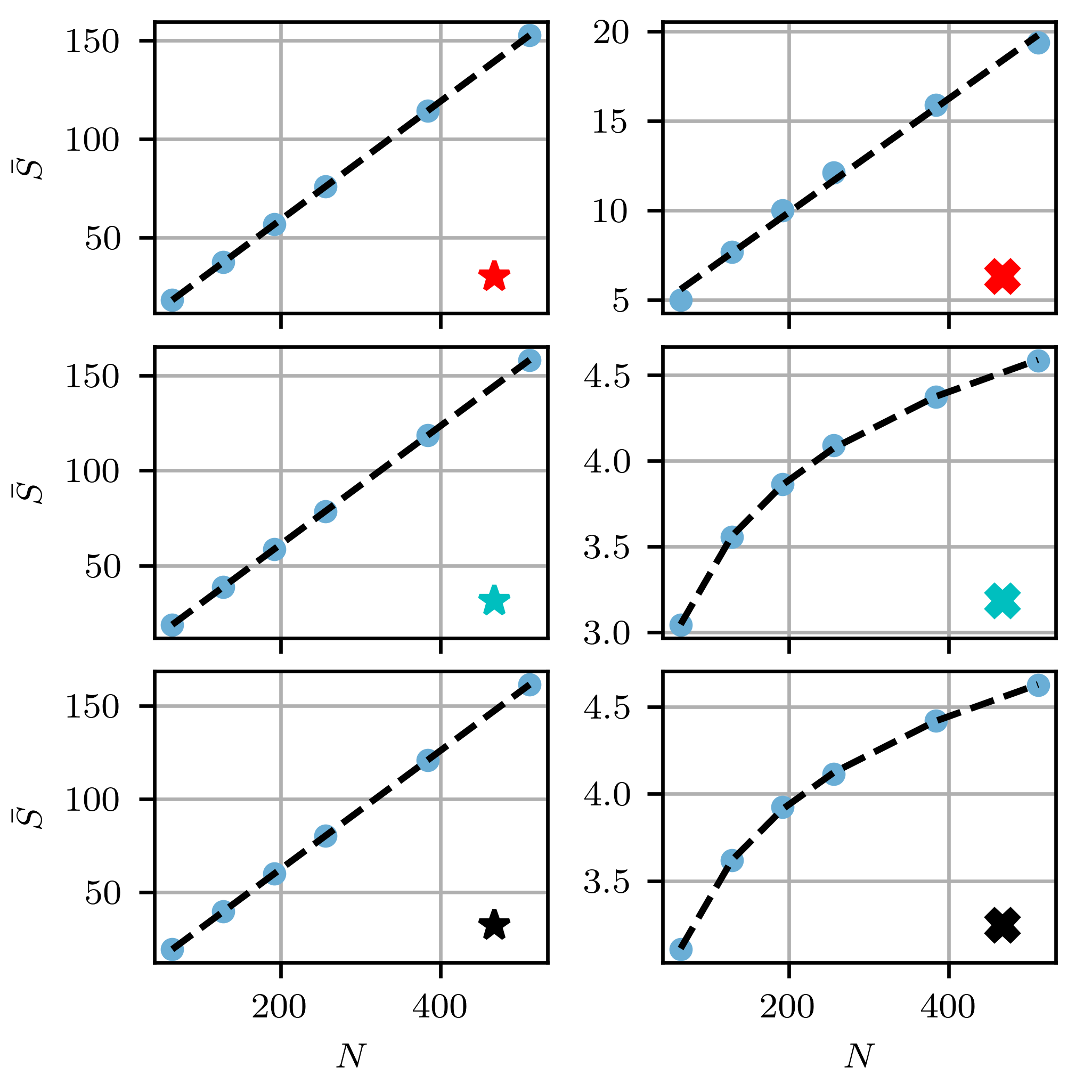} \label{subfig:rb-sfit}
}
\hfill 
\subfloat[][]{
\includegraphics[width=1.0\columnwidth]{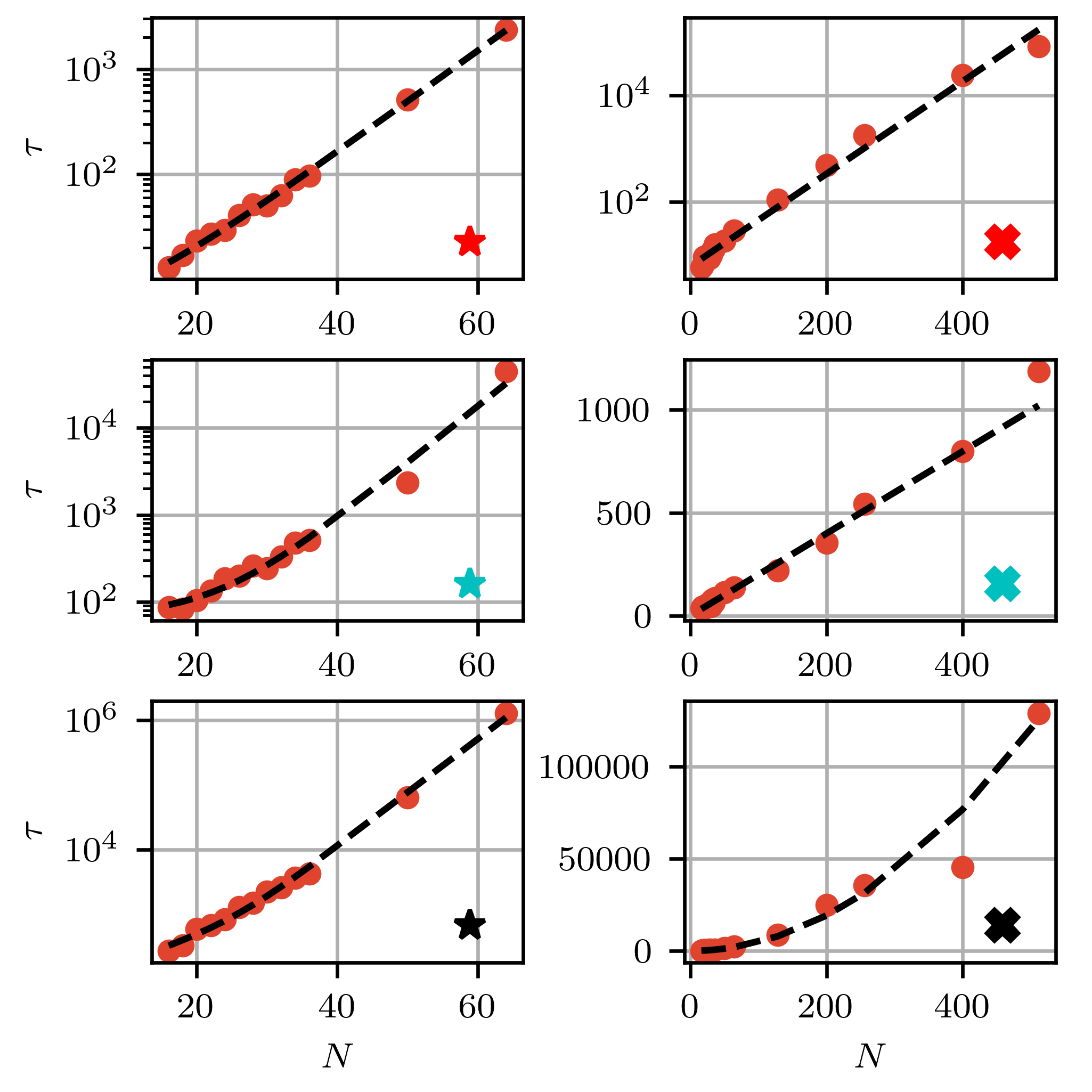}\label{subfig:rb-taufit}
}
\caption{ Here we examine six points across the random-basis phase diagram, as indicated by the matching marker and marker color in Fig. \ref{fig:rb-pd}. Left column of \ref{subfig:rb-sfit}, \ref{subfig:rb-taufit}: long-range measurements ($\alpha = 0$); right column of \ref{subfig:rb-sfit}, \ref{subfig:rb-taufit}: short-range measurements ($\alpha = 4$); from bottom to top: circuit density increasing with values $M_2/N = 0.0, 0.2, 0.5$. \ref{subfig:rb-sfit}: Steady state entanglement entropy vs system size. The data is accompanied by the best fit curve. For points in the long-range measurement regime and the maximally dense short-range limit, this is a linear function (volume-law entanglement), while for points with lower circuit density in the short-range measurement regime, this is a logarithmic function (critical entanglement). \ref{subfig:rb-taufit}: Purification timescales $\tau$ vs system size. The data is accompanied by the best fit curve. For points in the long-range measurement regime and the maximally dense short-range limit, this is an exponential function (non-purifying) which is plotted in semi-logarithmic scale, while for points with lower circuit density in the short-range measurement regime, this is a linear function (purifying). In the $M_2/N \rightarrow 0$ limit (bottom row), the data is first normalized by $N$ prior to fitting to account for the expected growth in $\tau$ by a factor of $N$ when measurement are performed strictly sequentially.}
\label{fig:rb-poi}
\end{figure*}

The color of the entanglement entropy phase diagram in Fig. \ref{fig:rb-pd} corresponds to the difference between the $R^2$ values, such that a positive value indicates a better linear fit (volume-law entangled) and a negative value indicates a better logarithmic fit (critical entanglement growth). Two regions appear across the parameter space, with dependence on both circuit density and measurement range. Generally, we observe that long-range measurements tend to produce volume-law entangled states, while sufficiently short-range measurements lead to sub-volume law entanglement growth. A simple intuition comes from the random-basis statistical mechanics mapping (Sec.~\ref{sec:methods}), where the entanglement entropy is the free-energy cost of imposing mixed identity/swap boundary conditions. In this picture, statistical weights are suppressed whenever the permutations entering a measurement block \(W_M\) or a Weingarten link disagree, so blocks that cross the identity-swap interface contribute an \(O(1)\) free-energy penalty and hence increase the entanglement.

Appendix~\ref{bipartition_scaling} summarizes a geometric estimate for how many such interface-straddling measurements occur on our simulation window \(t\sim N\). In the dense case (fixed measurement density \(M_2/N\)), the accumulated number of cut-crossing measurements up to \(t\sim N\) is \(O(N)\) for all \(\alpha\), so this estimate is consistent with volume-law entanglement throughout the \(\alpha\) range. In the asymptotically sparse case (\(M_2=O(1)\), i.e.\ \(M_2/N\to 0\)), the accumulated number of cut crossings yields an extensive (volume-law compatible) regime for \(\alpha<1\), but only subextensive growth for \(\alpha>1\) (with \(\log N\) at \(\alpha=2\) and \(O(1)\) for \(\alpha>2\)), consistent with under-volume-law scaling. We emphasize that this cut-crossing argument is purely geometric and although it applies equally to the Haar-averaged statistical mechanics model obtained by approximating random-basis MoC with random Haar unitaries, it should be viewed as a scaling heuristic rather than a determination of phase boundaries, which depend on the full set of statistical mechanics weights. With these caveats, the dense- versus sparse-regime expectations above are consistent with our numerical observations in both limits, as seen in Fig.~\ref{subfig:rb-sfit}.

A few examples of $\bar{S}(N)$ are plotted in Fig. \ref{subfig:rb-sfit} with the best fit function accompanying the data. Note that for each plot, the corresponding location in parameter space is indicated by the matching marker plotted over the phase diagram in Fig. \ref{fig:rb-pd}. Nearly identical volume-law entanglement growth appears for each $\alpha = 0$ point selected (long-range measurements), but when $\alpha = 4$, the form of entanglement growth is dependent on circuit density. At low circuit density, the entanglement growth is logarithmic. However, as the circuit density approaches its maximum value, the critical entanglement growth is superseded by volume-law entanglement.

\begin{figure}[!htb]
\centering
\includegraphics[width=1.0\columnwidth]{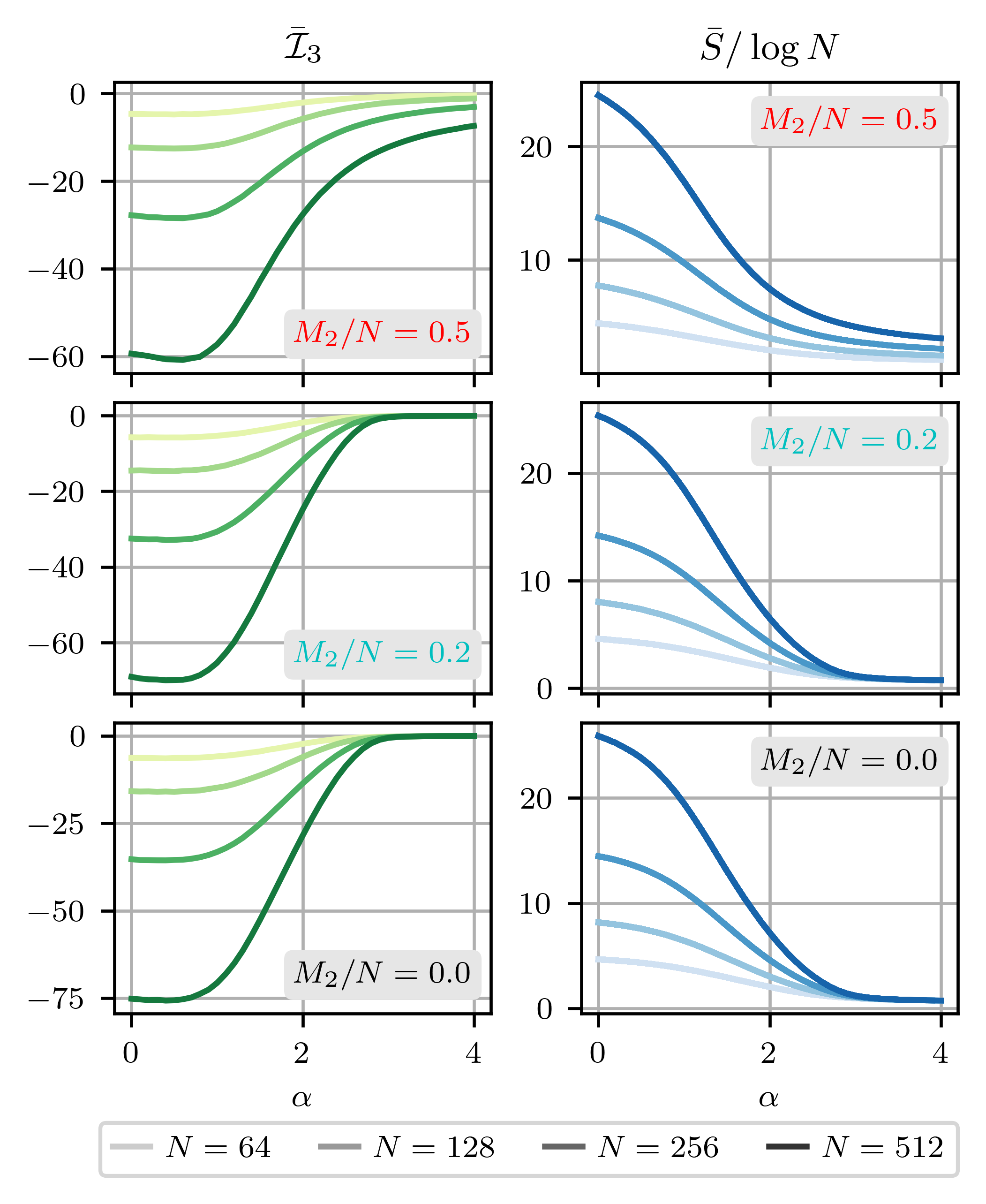}
\caption{Steady state values of tripartite mutual information and entanglement entropy vs interaction range parameter $\alpha$ for three values of circuit density $M_2/N$ and $N$ ranging from 64 (lightest curves) to 512 (darkest curves). Entropy values are weighted by $\log N$ to highlight the regions of critical entanglement growth for $M_2/N = 0.0, 0.2$, where the curves collapse on each other. The curves of $\bar{\mathcal{I}}_3$ similarly appear to collapse on each other in this region, with values near zero. For the densest circuits (top row), no collapse is observed.}
\label{fig:rb-crossover}
\end{figure}

In the right panels of Fig. \ref{fig:rb-crossover}, we examine the steady state entanglement entropy normalized by $\log N$ across horizontal cuts of the phase diagram for different system sizes. The curves collapse on each other in regions of logarithmic entanglement growth and remain separated in regions of volume-law entanglement. All results suggest the presence of an entanglement transition (i) in the measurement range $\alpha$ when the circuit crosses a threshold density, and (ii) in the circuit density $M_2/N$ in short-range circuits. The transition (i) is numerically found to be around $\alpha \sim 3$, and the transition (ii) $M_2/N \sim 0.4$. 

We stress that the numerical results support the analytical predictions and heuristic arguments about the existence of these different regimes, although the precise locations of the entanglement transitions may depend on the microscopics of the MoC. We find that however effective Hamiltonian derived in Sec.~\ref{ref:effectiveXYphase}, a long-range interacting XX chain, not only provides further insights on the entanglement transition between volume to sub-volume law scalings, but also on the transition point. 

For short-range interactions, the 1D XX chain is in a \(U(1)\)-symmetric, critical phase and does not exhibit spontaneous breaking of its continuous \(U(1)\) symmetry in the ground state. With sufficiently long-range interactions \(J_{ij}\propto |i-j|^{-\alpha}\), however, true long-range order and spontaneous \(U(1)\) symmetry breaking can occur below a critical decay exponent near \(\alpha_c\simeq 3\)~\cite{PhysRevLett.119.023001}. Remarkably, this is also where we observe the volume to sub-volume law transition in the random-basis MoC numerics in Fig.~\ref{fig:rb-pd}(a).

In fact, after the mapping, the swap (domain-wall insertion) operator becomes
\(\mathcal{S}_{n,A}=\prod_{i\in A}\sigma_i^x\) in the effective spin Hilbert space, so that for \(n=2\):
\begin{align}\label{eq:entropy_result}
    e^{-\tilde{S}_A^{(2)}}=
    \frac{\left\langle\!\left\langle\mathcal{I}\left|\prod_{i \in A} \sigma_i^x \right|\psi_{\mathrm{gs}}\right\rangle\!\right\rangle}
    {\left\langle\!\left\langle\mathcal{I}\middle|\psi_{\mathrm{gs}}\right\rangle\!\right\rangle},
\end{align}
where \(|\psi_{\mathrm{gs}}\rangle\!\rangle\) is the ground state of the effective long-range XX hamiltonian and \(|\mathcal{I}\rangle\!\rangle\) is the boundary product state fixed by the mapping~\cite{PhysRevLett.128.010604}.  In the \(U(1)\)-symmetric phase (\(\alpha\gtrsim \alpha_c\)), correlations are power-law and the domain-wall insertion \(\prod_{i\in A}\sigma_i^x\) has only an algebraic effect on the matrix element in Eq.~\eqref{eq:entropy_result}, implying sub-volume-law (critical) scaling of \(\tilde{S}_A^{(2)}\) with \(N\)~\cite{PhysRevLett.128.010604}. By contrast, in the \(U(1)\)-broken phase (\(\alpha\lesssim \alpha_c\)), the ground state exhibits true long-range order and the same insertion carries an extensive free-energy cost (equivalently, the numerator of Eq.~\eqref{eq:entropy_result} is exponentially suppressed in \(|A|\), the size of subsystem $A$), leading to volume-law entanglement scaling~\cite{PhysRevLett.128.010604}.

\subsubsection{Purification}\label{sec:rb-purification}

In the lower right panel of Fig. \ref{fig:rb-pd}, we examine how the purification timescale $\tau$ scales with system size by comparing $R^2$ values when the $\tau$ versus $N$ data fit to an exponential and a linear function. Dark red indicates a preference for the exponential fit and suggests that purification does not occur, while light yellow indicates a preference for linear fit and suggests that purification will occur. Two main regimes appear: when measurements are long-range, $\tau$ grows exponentially with system size, see Fig. \ref{subfig:rb-taufit}. We expect the ancilla qubit $a$ to remain entangled with the system during the circuit evolution for this region of parameter space. Hence, the circuit does not purify in the phase of volume-law entanglement entropy. 

Transitioning to short-range measurements, $\tau$ instead scales linearly with system size. We observe the circuit evolution to disentangle $a$ from the system in this region as the system purifies in this sub-volume law entangled phase. Once again, a short-range, dense circuit is the exception: the exponential growth of $\tau$ with $N$ persists, and the circuit does not purify (top right corner of Fig. \ref{subfig:rb-taufit}), which is consistent with the volume-law entanglement entropy scaling, as discussed in the previous section \ref{entanglement_growth1}. Therefore, the entanglement entropy transitions in the measurement range and the circuit density are also visible in the purification results.  

The transition with the measurement range can be elucidated with the idea of measurement frustration \cite{PhysRevX.11.011030}. In an ensemble where all measurements commute with each other, the state will purify into a simultaneous eigenstate of the measurements leading to an area-law entanglement entropy. However, by the uncertainty principle, non-commuting measurements cannot be all known at the same time, causing a measurement frustration. In a random-basis circuit with many layers, the fraction of measurements that commute with each other is dictated by $\alpha$. This is simply because it is more likely to measure the same two qubits sequentially in different bases in a local circuit than in a nonlocal circuit. Concretely, when $\alpha = 2$, the probability that a given measurement has a range 1 is $1/\zeta(2) =6/\pi^{2} = 0.61$, and when $\alpha = 4$ it is $1/\zeta(4) =90/\pi^4 = 0.92$. Thus there are more commuting measurements in the local than in the nonlocal circuit, leading to a measurement frustration as the measurement range increases.   

In the limit of $M_2/N \rightarrow 0$ with short-range measurements (lowest right plot in Fig. \ref{subfig:rb-taufit}), $\tau$ is seen to grow quadratically with $N$. This extra factor of $N$ arises because the number of measurements in the $M_2/N \rightarrow 0$ limit remains fixed as $N$ increases, so that no two measurements are made concurrently. This constraint in space leads to an additional factor of $N$ in the purification timescale $\tau$. 
Therefore, we observe a dynamical critical exponent of $z=1$ throughout the purifying section of the phase diagram.

\subsubsection{Scrambling and Non-Scrambling Regimes}\label{sec:rb-scrambling}
The top left plot in Fig. \ref{fig:rb-pd} displays the steady state value of the tripartite mutual information $\bar{\mathcal{I}}_3$. Two regions emerge closely following the transition boundaries found in previous sections. In a long-range measurement circuit with small $\alpha$, the tripartite mutual information is negative with a large magnitude, which is indicative of information scrambling. As the measurements become shorter range, the amplitude of $\bar{\mathcal{I}}_3$ decreases, albeit it remains negative. In the left column of Fig. \ref{fig:rb-crossover}, we examine the shift in scrambling behavior versus $\alpha$ for different system sizes. These horizontal cuts across the phase diagram examines whether the transitions found in previous sections can also be probed by the TMI. 

In the long-range regime, $\bar{\mathcal{I}}_3$ becomes increasingly negative with larger systems, evidencing robust information scrambling. We label this region of parameter space the \textit{scrambling regime}. For not maximally dense circuit densities, the curves collapse in the short-range regime. Although this is suggestive of $\mathcal{I}_3 \rightarrow 0$ in the thermodynamic limit, we do not find the data in logarithmic scale conclusive to claim a transition in the thermodynamic limit. This might be due to the temporal fluctuations in $\mathcal{I}_3$ even in its steady state, see Appendix~\ref{appendix:sim}. 
However due to the observation that $\mathcal{I}_3$ behaviors are distinctly different in long- and short-range circuits when the circuit density is sparse enough, we conclude that the TMI is susceptible to the transition in the measurement range, and can be used to probe it. In the dense circuit limit $M_2/N=0.5$, the curves remain separated with $\bar{\mathcal{I}}_3 < 0$, and the magnitude of $\bar{\mathcal{I}}_3$ increases with $N$ for all values of $\alpha$. Therefore, TMI is also susceptible to the transition in the circuit density in the short-range measurement circuits. 

\begin{figure}[!htb]
\centering
\includegraphics[width=1.0\columnwidth]{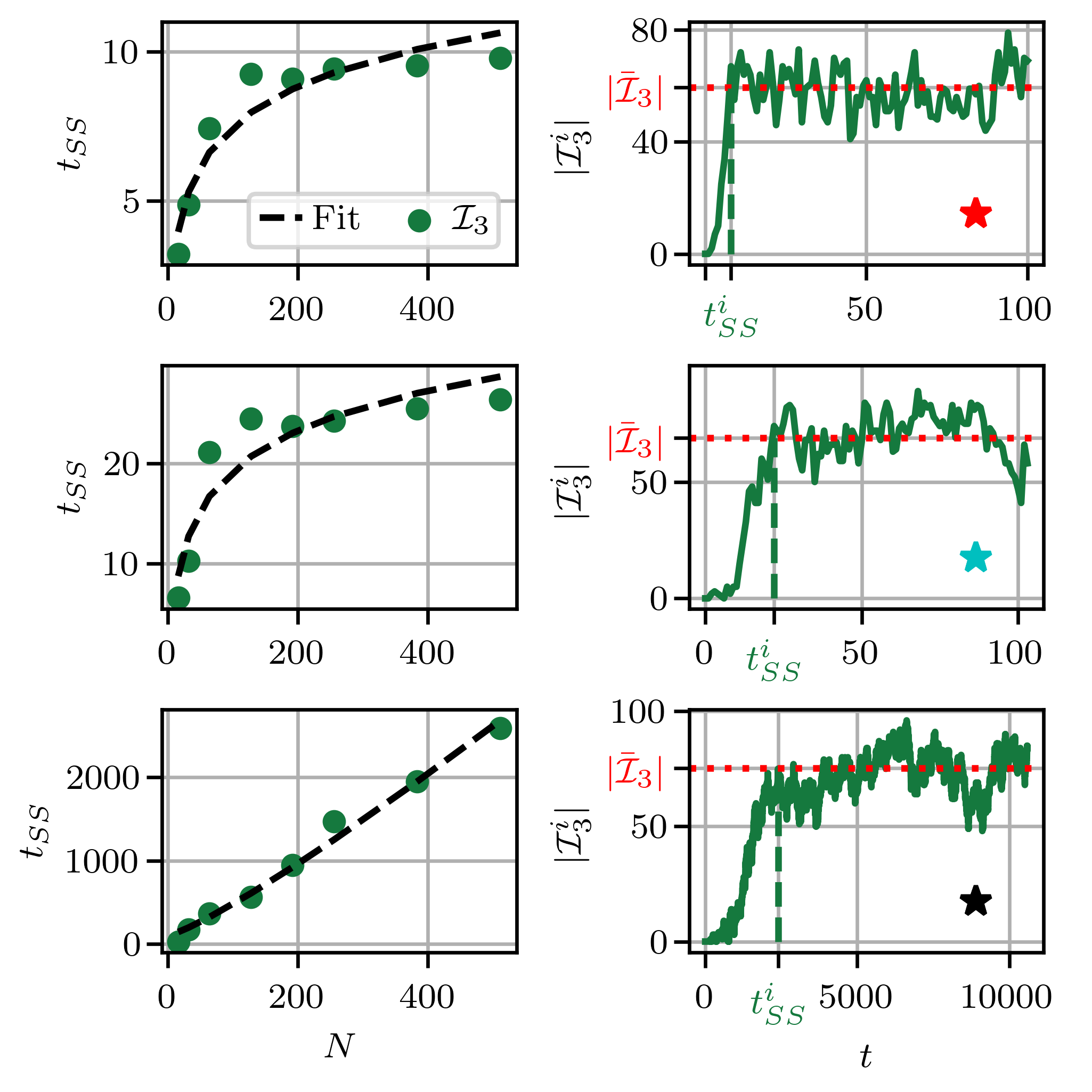}
\caption{ Time to steady state of tripartite mutual information for the random-basis circuit setup. Only $\alpha = 0$ is considered (long-range measurement regime, where $\mathcal{I}_3 \ll 0$), for three different circuit densities: $M_2/N = 0$ (bottom), $M_2/N = 0.2$ (middle), and $M_2/N = 0.5$ (top). The left plots examine the trajectory averaged values of $t_{SS}$ vs system size, with the accompanying best fit curve: logarithmic for $M_2 / N > 0$, and $N \log N$ for the $M_2 / N = 0$ limit. The right plots give a single trajectory example of $|\mathcal{I}^i_3|$ for $N = 512$, with the corresponding steady state value plotted horizontally in red with the dotted line and the computed value of $t_{SS}^{i}$ plotted vertically with the dashed line. }
\label{fig:rb-tSS}
\end{figure}

In the scrambling regime, an interesting question arises regarding the number of circuit layers required to achieve steady state and how this quantity scales with system size. This question relates to the speed of scrambling. We define the time to steady state $t^{i}_{SS}$ as the number of circuit layers required for a particular circuit trajectory $i$ to be within 1\% of the steady state value for a quantity of interest. In the left panels of Fig. \ref{fig:rb-tSS}, we plot $t_{SS}$ averaged over circuit trajectories for TMI; this quantity can be viewed as the ``time to scrambling.'' Three values of $M_2 / N$ are considered in the long-range measurement regime ($\alpha = 0$). In the sparse circuit limit of $M_2/N \rightarrow 0$, the circuit is constructed to have only one measurement per layer even as $N$ increases. Structurally, this regime can be compared to random quantum circuits considered by \cite{harrow_random_2009}, which are known to scramble information in $\mathcal{O}(N\log N)$ time \cite{sekino_fast_2008}. Our setup replaces the two-qubit unitary interactions with two-qubit projective measurements, yet we posit that the time to scrambling in this limit also grows as $\mathcal{O}(N\log N)$. When the circuit density is finite ($M_2 / N > 0$), the number of measurements per layer increases with $N$, such that the fraction of qubits involved in measurements within a given layer remains constant. Intuitively, stacking the measurements into a ``parallel processing'' architecture with fewer layers will decrease the number of layers required to reach steady state. In Ref.~\cite{hayden_black_2007}, Hayden and Preskill considered a similar adjustment to the random unitary circuit to produce a fast scrambling system. Here, we similarly find that measurement-only circuits with parallel measurements also saturate the fast scrambling limit, with $t_{SS}$ growing as $\mathcal{O}(\log N)$.

\begin{figure*}[!htb]
\centering
\subfloat[][]{
\includegraphics[width=1.0\columnwidth]{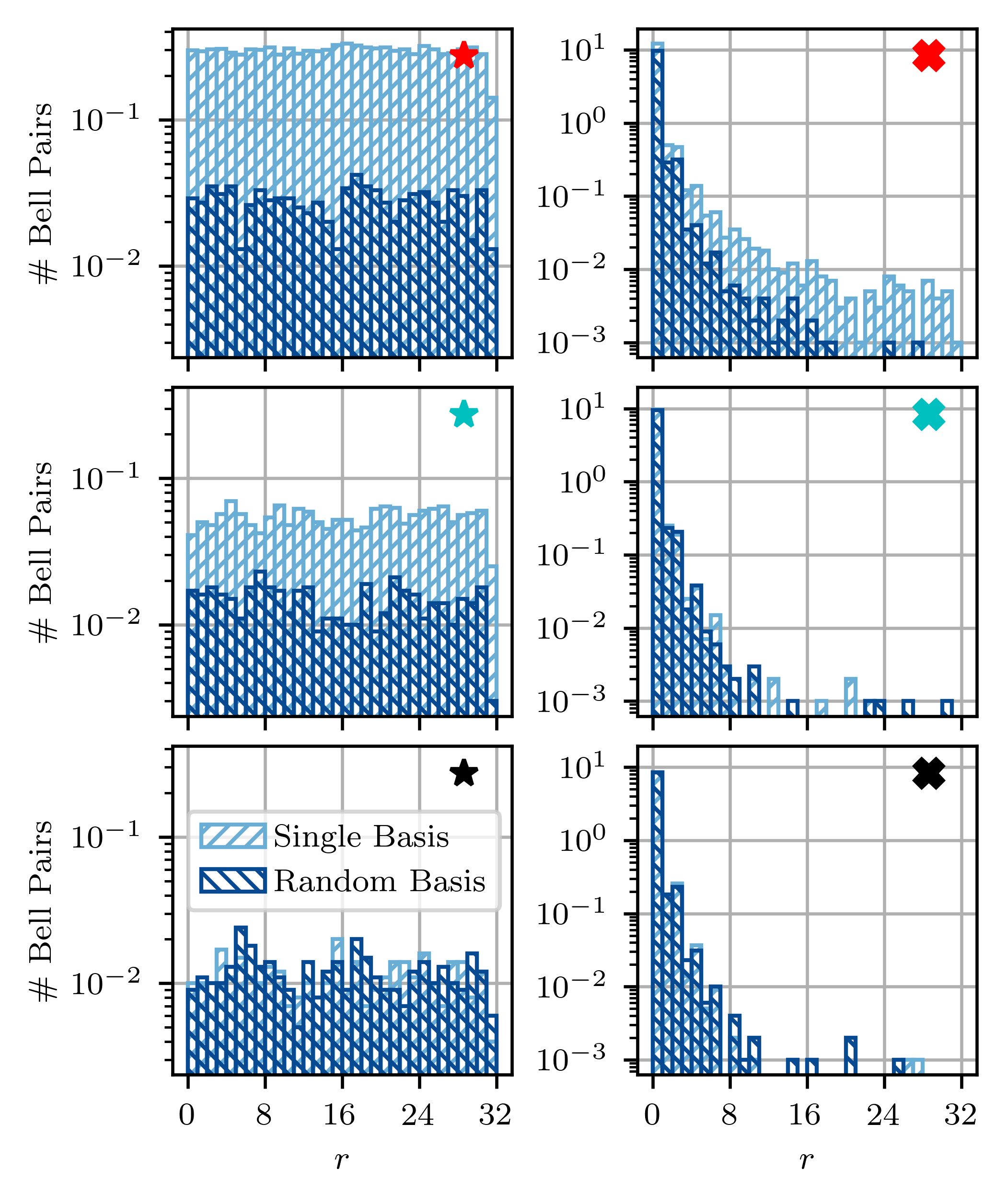}\label{subfig:bell}
}
\hfill 
\subfloat[][]{
\includegraphics[width=1.0\columnwidth]{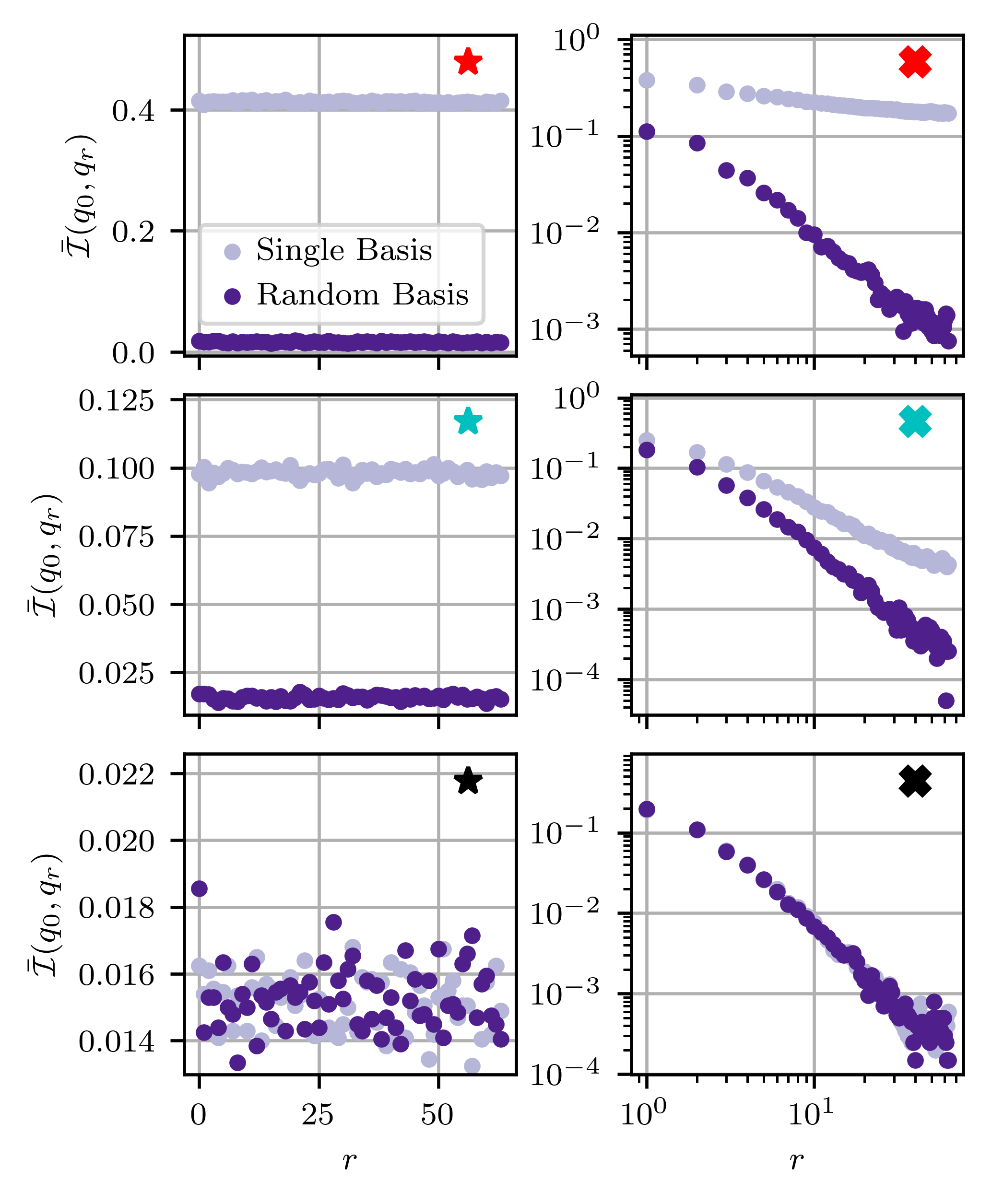}\label{subfig:Ivsr}
}
\caption{ Here we consider six points across the parameter space, as indicated by the matching marker and marker color in Fig. \ref{fig:rb-pd}. Both single-basis (light blue / purple) and random-basis (dark blue / purple) scenarios are considered and contrasted. Note that in the regime of the bottom plots ($M_2/N = 0$), the single and random-basis cases are identical. \ref{subfig:bell}: Average number of Bell pairs observed in the final state of 1000 circuit trajectories. Data is arranged according to the distance between the maximally entangled qubits, with $1 \leq r \leq N/2$ ($N = 64$). Number of Bell pairs increases with increasing circuit density, particularly for the single-basis case. Distribution over $r$ is uniform for $\alpha = 0$ (long-range measurements), but peaks at $r=1$ for $\alpha = 4$ (short-range measurements). \ref{subfig:Ivsr}: Steady state mutual information between qubits at position 0 and $r$ for $1 \leq r \leq N/2$ ($N = 128$). For $\alpha = 0$ (long-range measurements), no decay is observed with increasing $r$. $\bar{\mathcal{I}}$ is generally larger for the single-basis case, and this separation increases with increasing circuit density. For $\alpha = 4$ (short-range measurements), there is power-law decay of $\bar{\mathcal{I}}$ with $r$ for the random-basis case, and the specific power depends on circuit density. For larger densities, the single-basis data initially exhibits power law decay, but $\bar{\mathcal{I}}$ then stabilizes to a non-zero value.}
\label{fig:bell}
\end{figure*}

\subsubsection{Mutual information}\label{sec:rb-mutualinfo}
The top right plot in Fig. \ref{fig:rb-pd} displays the steady state value of the mutual information $\bar{\mathcal{I}}$ between two maximally distant qubits in the system. The value of $\bar{\mathcal{I}}$ remains small throughout the phase diagram, with marginally elevated values in the long-range measurement regime (small $\alpha$). Fig. \ref{subfig:Ivsr} examines mutual information in more detail by plotting it between two qubits in the system separated by a distance $r$ where the random-basis results are provided in dark purple. In the long-range measurement limit ($\alpha = 0$), the mutual information is constant with space $r$, leading to the observed nonzero but small value for $r = N/2$ in Fig. \ref{fig:rb-pd}. We can identify this constant but small value in $\bar{\mathcal{I}}$ with the presence of a suppressed long-range entanglement structure. This is because mutual information is sensitive to the number of Bell clusters in a state \cite{PhysRevB.102.094204}. 

Fig.~\ref{subfig:bell} plots the number of Bell \textit{pairs} detected in different ranges $r$ (dark blue denotes the random-basis circuits). To find this, we first compute the entanglement entropy between all possible pairs of qubits in a small system ($N = 64$) after reaching steady state. If the qubits residing at sites $i$ and $j$ are maximally entangled (and therefore equivalent to a Bell pair up to single qubit rotations), then the entanglement entropy $S_{i,j}$ will equal 2 and the entanglement entropy of $i$ ($j$) with all qubits $k \neq j$ ($k \neq i$) will vanish: $S_{i, k \neq j} = S_{k \neq i, j} = 0$. We aggregate all such Bell pairs that are observed for 1000 circuits, organizing the observed pairs according to the distance $r$ between qubits at sites $i$ and $j$. Fig. \ref{fig:bell} then presents the average number of pairs found at each $r$ from points across the phase diagram. The uniform distribution over the space suggests that it is equally likely to find a Bell pair at any range in a long-range measurement circuit. While this is consistent with the expectation of long-range entanglement, the overall amplitudes are small, which implies the suppression of this long-range entanglement. This is reasonable, as this phase is volume-law entangled, non-purifying and scrambling.

\begin{figure*}[!htb]
\centering
\subfloat[][]{
\includegraphics[width=0.75\columnwidth]{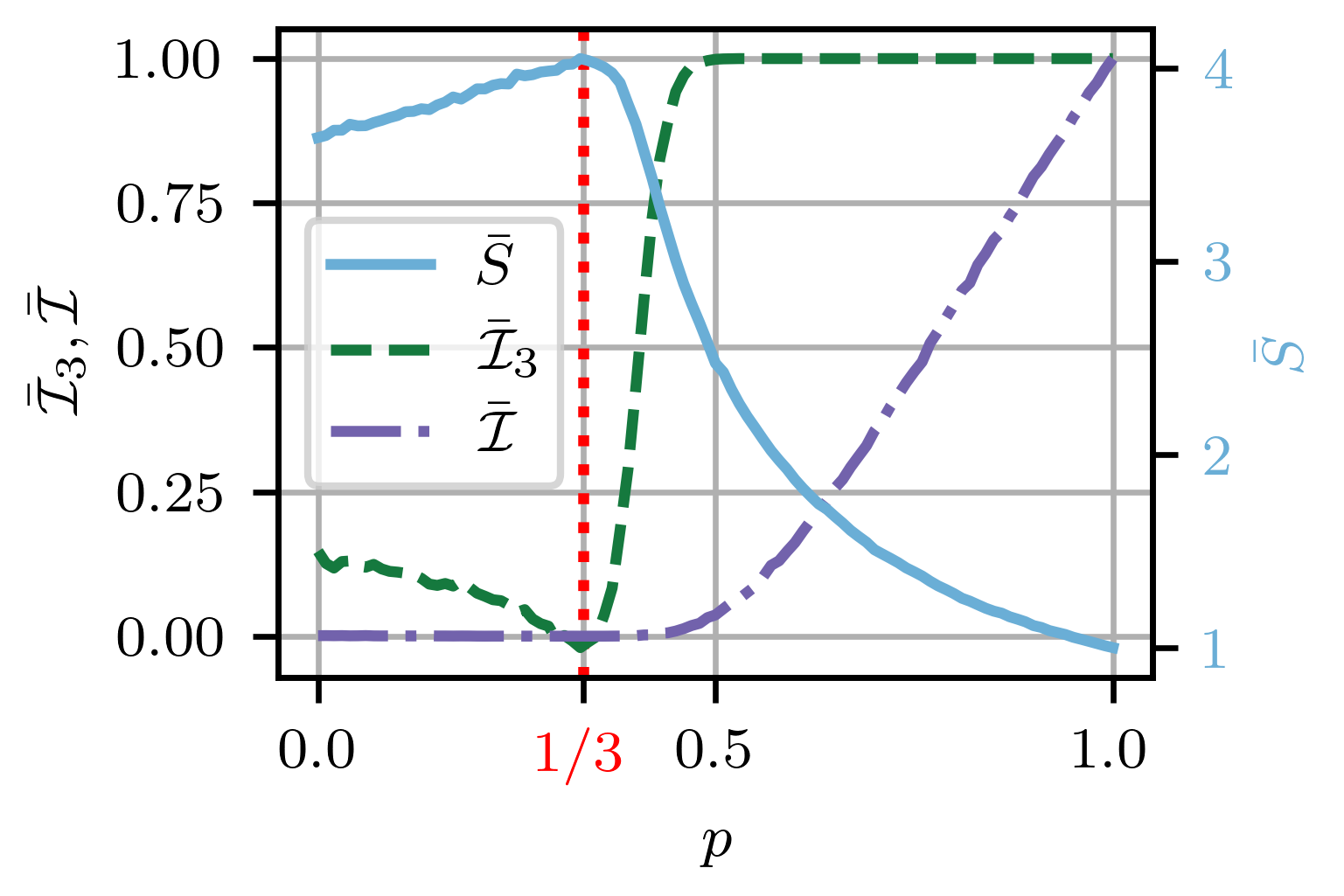}\label{subfig:xxz-p}
}
\hfill 
\subfloat[][]{
\includegraphics[width=1.25\columnwidth]{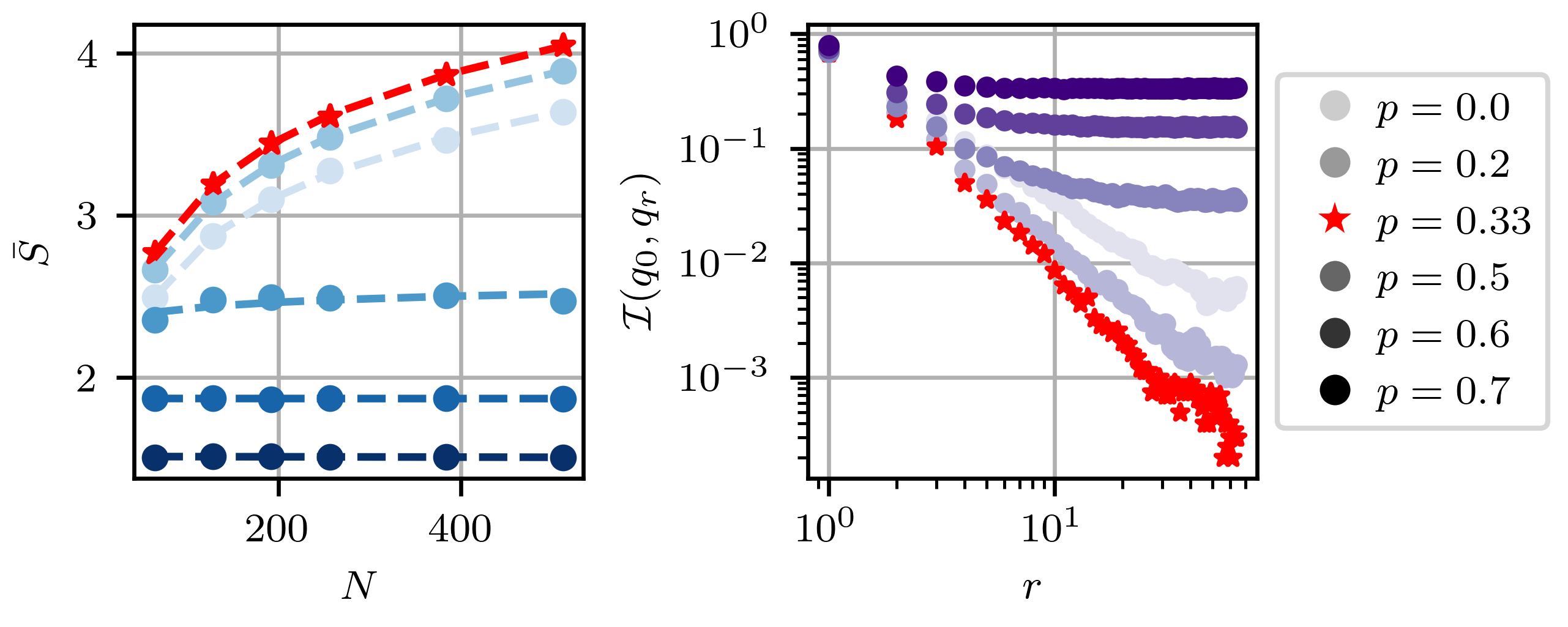}\label{subfig:xxz-Sfit-Idecay}
}
\caption{ Projective XXZ model. Steady state values for TMI, mutual information, and entanglement entropy as a function of $p$ for a system of size $N = 512$. The role of long-range order in the Ising-like phase, i.e.,~a circuit dominated by $Z_iZ_j$, is played by the mutual information between two far away spins which is vanishing in the XY-like phase, i.e.,~a circuit with equally dominated by $X_iX_j$ and $Y_iY_j$. Note that the entanglement entropy is at its maximum at the transition point. This is also the only scrambling point in the phase space. 
\ref{subfig:xxz-Sfit-Idecay}: The entanglement entropy with respect to system size (left) and the mutual information between qubit $q_0$ and a qubit a distance $r$ away is plotted as a function of $r$ for a system of size $N = 128$. The results show that XY-like phase is a critical region with sub-volume law entanglement entropy and power-law decaying mutual information in distance, whereas the Ising-like phase has area-law entanglement, and finite and constant mutual information. }
\label{fig:xxz}
\end{figure*}
As the measurement-range decreases, a power-law decay in mutual information with $r$ emerges in Fig.~\ref{subfig:Ivsr}, leading to  vanishingly small value of mutual information on the right side of the plot in Fig. \ref{fig:rb-pd}. Importantly, the power-law decay in the mutual information is consistent with the critical sub-volume law scaling of the entanglement entropy and a purifying phase seen in the short-range and sufficiently sparse circuits. We find that Bell pairs are also more likely to occur in short ranges. Let us note that the mutual information does not see the transition in the circuit density unlike other observables. Hence, this leads to a phase in the dense and short-range circuits that is non-purifying, scrambling and volume-law entangled, with no long-range entanglement.  

\subsubsection{Projective XXZ Model}\label{XXZ section}

We emphasize that all random-basis circuits discussed can be considered as \textit{projective Heisenberg models} in the steady-state, meaning that these circuits consist of sequentially applied measurements in the equally weighted bases,  
similar to the projective Ising model introduced in \cite{PhysRevB.102.094204}. The projective Heisenberg models also effectively generate the ground state properties of the XX chain with power-law decaying interactions, as we found in Sec.~\ref{ref:effectiveXYphase}. 
In this section, inspired from this finding, we introduce a circuit modification to realize a projective and locally connected XXZ model.  Concretely, the parameter $p$ weights the probability of selecting $Z_{i} Z_{j}$ so that the probability of selecting $X_{i} X_{j}$ or $Y_{i} Y_{j}$ is equal to $(1 - p)/2$. We fix a sparse circuit with $M_2/N \rightarrow 0$ (corresponding to one measurement per layer) and focus on the $\alpha \rightarrow \infty$ case (nearest-neighbor measurements only) in order to simplify the parameter space. Fig. \ref{subfig:xxz-p} plots the observed steady state values for entanglement entropy, mutual information, and tripartite mutual information upon sweeping the probability $p$. 

The point $p = 1/3$ in the phase diagram, as discussed in previous sections, corresponds to a critical gapless phase in the effective Hamiltonian, as well as, to the transition point from an area-law phase at $p> 1/3$. The entanglement entropy is maximized at this point in the phase diagram and exhibits a logarithmic increase with the system size \cite{PhysRevB.102.094204}, (left panel in Fig.~\ref{subfig:xxz-Sfit-Idecay}). The point $p=1/3$ is the only point where the circuit has SU(2) symmetry owing to the isotropic measurement rates in all basis. Together with the translational symmetry of the circuit, due to periodic boundary conditions, Lieb-Schultz-Matthis theorem forbids an area-law scaling in entanglement entropy \cite{nahum_entanglement_2020,majidy_critical_2023}. Further evidence for the criticality at this point is found in the mutual information, which decays in space with a power-law exponent $\kappa \sim 1.8$ (in the case with local measurements). We also determine that this is the only point in the entire projective-XXZ circuit that exhibits scrambling with negative TMI.

The region $p < 1/3$, where the circuit is equally dominated by $X_iX_j$ and $Y_iY_j$ measurements, exhibits sub-volume law in the entanglement entropy and power-law decay in space in the mutual information, Fig.~\ref{subfig:xxz-Sfit-Idecay}. We note a continuous decrease in the power-law $\mathcal{I}$ decay exponent as $p$ approaches a circuit free of $Z$ measurements, $p=0$. These results suggest that the modification to the circuit in the region $p < 1/3$ does not change the nature of the ground state in the effective Hamiltonian, remaining in the XY phase. Consistent with these observations, no scrambling is expected in this region \cite{PhysRevLett.123.140602}, which is confirmed with positive time-averaged TMI $\bar{\mathcal{I}}_3$.

The region $p>1/3$ where the circuit is dominated by measurements in the $Z$ basis, exhibits area law entanglement entropy which vanishes as $p \rightarrow 1$. This is expected since in a circuit with only $ZZ$ measurements no entanglement can build up. The area-law entanglement entropy was also found in the projective Ising model \cite{PhysRevB.102.094204}, suggesting that the region $p>1/3$ is an Ising-like phase. 
Consistently with an area-law behavior, we do not observe scrambling, and the TMI in its steady-state exhibits a sharp transition to its maximum value. Mutual information also follows suit and remains nonzero in long distances. Since it is exactly zero at $p<1/3$ and nonzero at $p>1/3$, mutual information acts like an order parameter \cite{PhysRevB.102.094204}. Particularly, its nonzero and large values signify the potential of this phase for possessing long-range entanglement when $p < 1$. These results also imply that the effective Hamiltonian of the circuit in the steady state goes through a transition once $p > 1/3$ holds, suggesting that $p$ parameter must induce a term competing with the flip-flop terms, mirroring the competition in the circuit itself.

\subsection{Single-basis measurement model}
\begin{figure}[!htb]
\centering
\includegraphics[width=1.0\columnwidth]{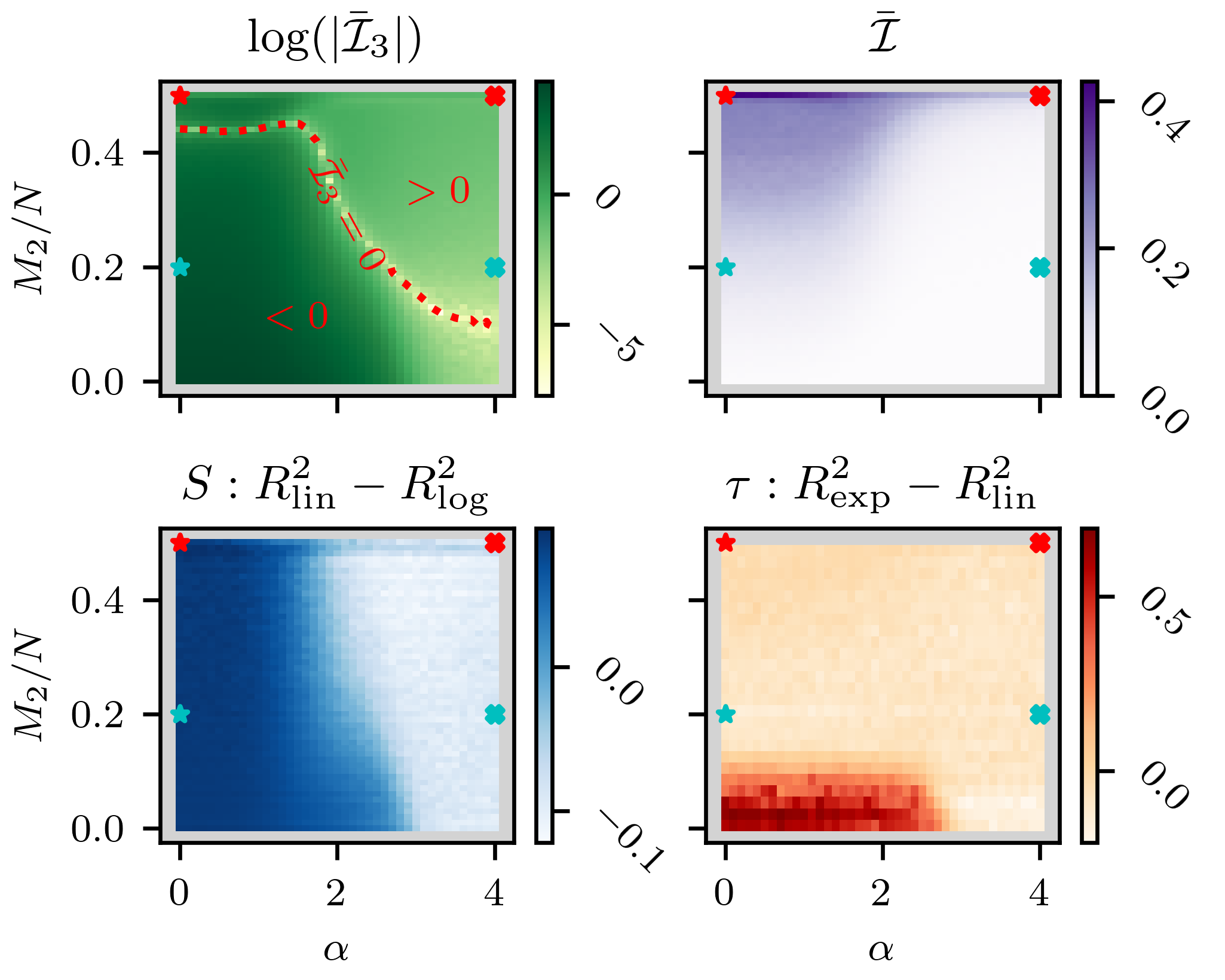}
\caption{Phase diagrams for single-basis measurement scheme. Probabilistic measurement range decreases from left to right, and circuit density increases from bottom to top. Top left plot: steady state log of the absolute value of tripartite mutual information. The sign of $\bar{\mathcal{I}}_3$ transitions from negative in the lower left to positive, as indicated by the overlaid text and zero contour line. Top right plot: steady state mutual information between two maximally distant qubits. Bottom left plot: difference between fitting $R^2$ values when steady state bipartite entropy data vs system size $N$ is fit to a linear function and a logarithmic function. Positive regions (dark blue) indicate volume law entanglement ($R^2_{\text{lin}} > R^2_{\text{log}}$) while negative regions (light blue) indicate sub-volume law entanglement ($R^2_{\text{lin}} < R^2_{\text{log}}$). Bottom right plot: difference between fitting $R^2$ values when purification timescale $\tau$ vs system size $N$ is fit to an exponential function and a linear function. Positive regions (dark red) indicate non-purifying ($R^2_{\text{exp}} > R^2_{\text{lin}}$) while negative regions (yellow) indicate purifying ($R^2_{\text{exp}} < R^2_{\text{lin}}$). The $M_2 / N = 0$ limit (one measurement per layer regardless of $N$) is treated slightly differently: to account for the expected additional factor of $N$ in time, the values of $\tau$ are divided by $N$ prior to fitting. Note that the maximally dense ($M_2/N$) results are omitted from this particular figure due to the fact that neither a linear or exponential function providing a good fit for $\tau(N)$}
\label{fig:sb-pd}
\end{figure}

In the single-basis case, all measurements within one circuit layer are constrained to be in the same basis. This change reduces the non-commutativity of the randomly generated circuit; however, 
also makes it more amenable to experimental implementation. We repeat the numerical simulation of 1000 circuits across the parameter space. Fig. \ref{fig:sb-pd} provides the resulting phase diagrams. In the following sections, we omit detailed discussion of the sparse $M_2 / N \rightarrow 0$ limit, which is identical to the $M_2 / N \rightarrow 0$ results for the random-basis case. 

\begin{table}[!htb]
    \begin{center}
    \caption{Summary of findings for single-basis MoCs}
    \label{tab:sb}
    %\begin{tabular}{|>{\bfseries}c|>{\centering\arraybackslash}m{1.2cm} | >{\centering\arraybackslash}m{1.2cm}|>{\centering\arraybackslash}m{1.3cm}|>{\centering\arraybackslash}m{1.3cm}|}
    \begin{tabular}{|c|c|c|c|c|}
        \hline \rule{0pt}{12pt}
        & \textbf{$\mathcal{I}_3$} & \textbf{$\mathcal{I}$} & \textbf{$S$} & \textbf{$\tau$} \\[2pt]
        \hline
        \makecell{\textbf{short range} \\  \textbf{sparsest} \\ $\alpha \sim 4, M_2/N \sim 0$ } & $\lesssim 0$ & $=0$ & $\mathcal{O}(\log N)$ & $\mathcal{O}(N)$ \\
        \hline 
        \makecell{\textbf{short range} \\ \textbf{sparse layers} \\ $\alpha \sim 4, M_2/N > 0 $} & $\gtrsim 0$ & $=0$ & $\mathcal{O}(\log N)$ & $\mathcal{O}(N)$ \\
        \hline 
        \makecell{\textbf{short range} \\ \textbf{dense layers} \\ $\alpha \sim 4, M_2/N \lesssim 0.5 $} & $\gtrsim 0$ & $\gg 0$ & $\mathcal{O}(\log N)$ & $\mathcal{O}(1)$ \\
        \hline
        \makecell{\textbf{long range} \\ \textbf{sparsest layers} \\ $\alpha \sim 0, M_2/N \sim 0$} & $<0$ & $\neq 0$ & $\mathcal{O}(N)$ & $\mathcal{O}(e^N)$ \\
        \hline
        \makecell{\textbf{long range} \\ \textbf{sparse layers} \\ $\alpha \sim 0, M_2/N > 0$} & $<0$ & $\gg 0$ & $\mathcal{O}(N)$ & $\mathcal{O}(N)$\\
        \hline
        \makecell{\textbf{long range} \\ \textbf{dense layers} \\ $\alpha \sim 0, M_2/N \lesssim 0.5$} & $>0$ & $\gg 0$ & $\mathcal{O}(N)$ & $\mathcal{O}(1)$\\
        \hline
        \bottomrule
    \end{tabular}
    \end{center}
\end{table}

\subsubsection{Phases of Entanglement Growth}
\begin{figure*}[!htb]
\centering
\subfloat[][]{
\includegraphics[width=1.0\columnwidth]{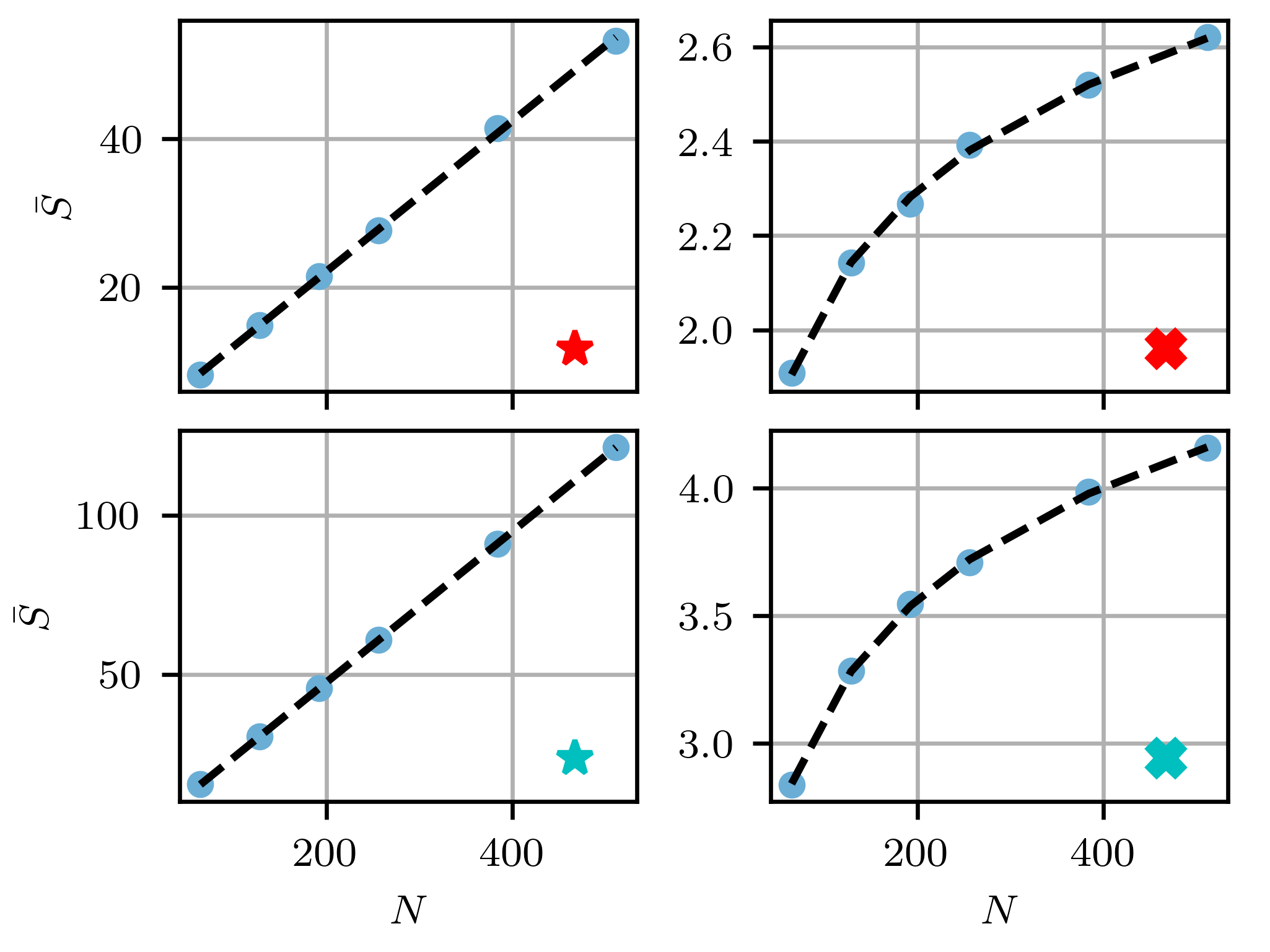}\label{subfig:sb-sfit}
}
\hfill 
\subfloat[][]{
\includegraphics[width=1.0\columnwidth]{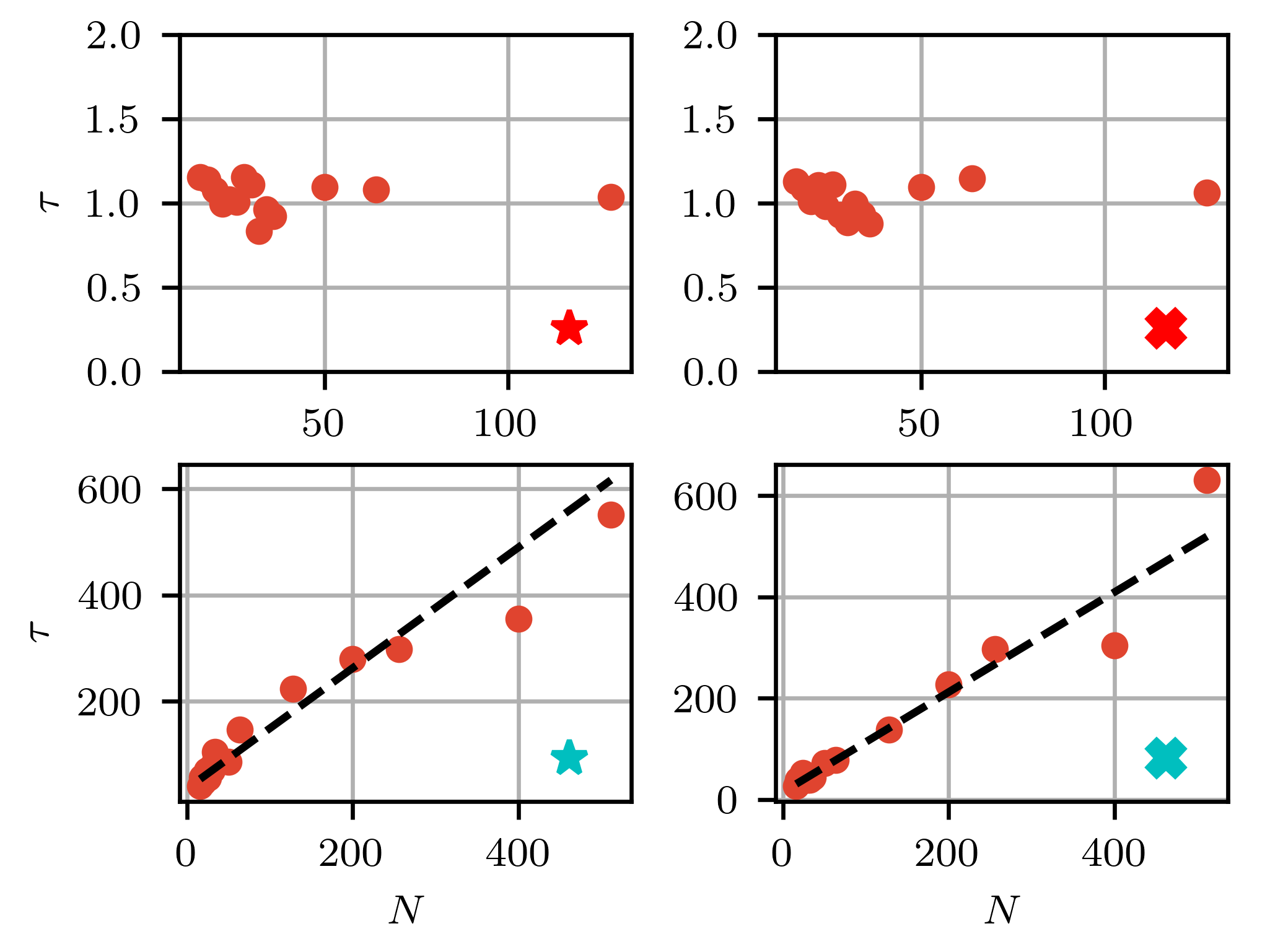}\label{subfig:sb-taufit}
}
\caption{Here we examine four points across the single-basis phase diagram, as indicated by the matching marker and marker color in Fig. \ref{fig:sb-pd}. Left column of \ref{subfig:sb-sfit}, \ref{subfig:sb-taufit}: long-range measurements ($\alpha = 0$); right column of \ref{subfig:sb-sfit}, \ref{subfig:sb-taufit}: short-range measurements ($\alpha = 4$); from bottom to top: circuit density increasing with values $M_2/N = 0.2, 0.5$. \ref{subfig:sb-sfit}: Steady state entanglement entropy vs system size. The data is accompanied by the best fit curve. For points in the long-range measurement regime, this is a linear function (volume law entanglement), while for points in the short-range measurement regime, this is a logarithmic function (critical entanglement). \ref{subfig:sb-taufit}: Purification timescale $\tau$ vs system size. The data is accompanied by the best fit curve. For moderate circuit density across all measurement ranges, this is a linear function (purifying). When circuit density is maximum (top row), purification occurs exceptionally fast and $\tau$ is independent of $N$.}
\label{fig:sb-poi}
\end{figure*}

\begin{figure}[!htb]
\centering

\includegraphics[width=1.0\columnwidth]{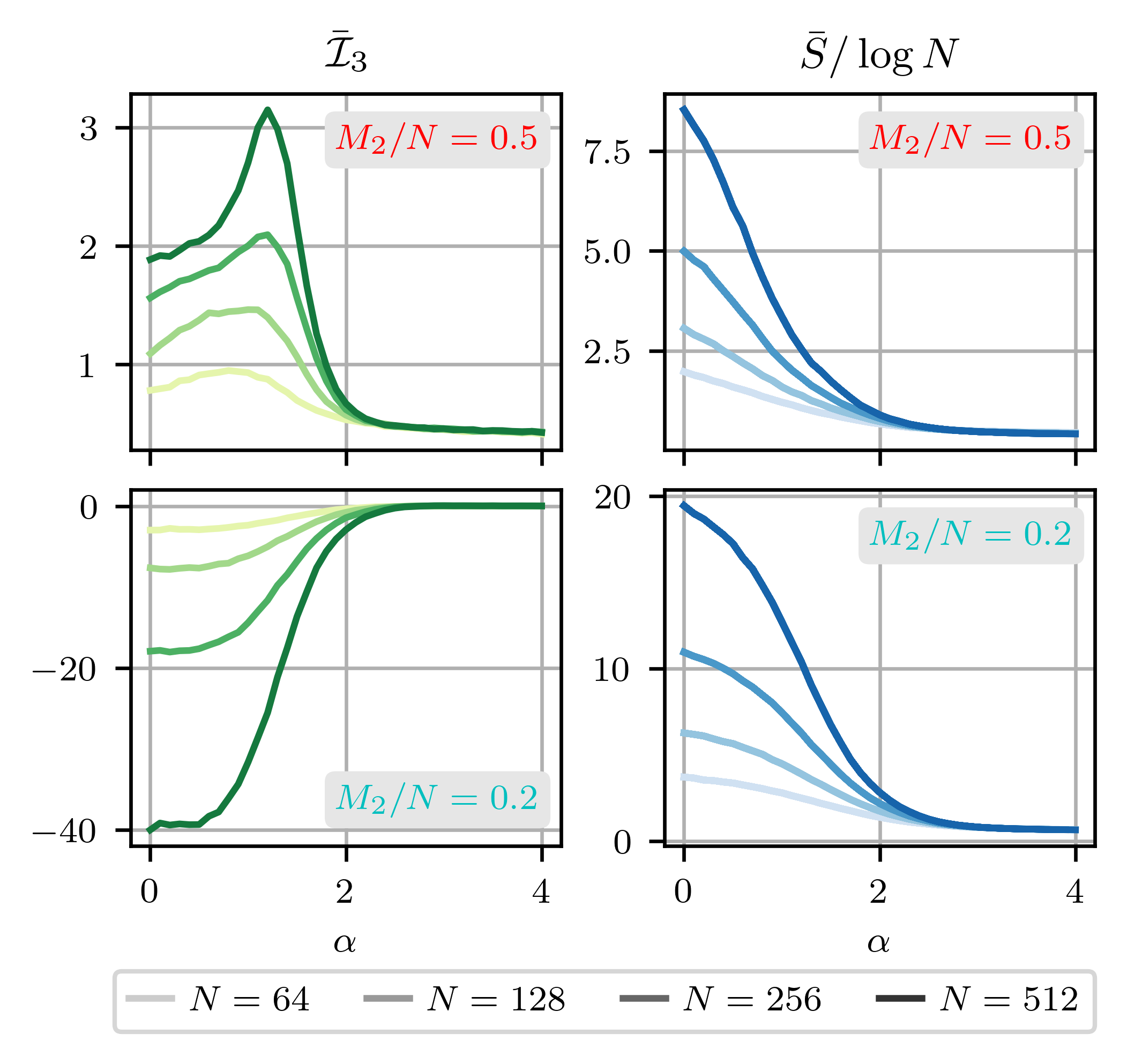}
\caption{Steady state values of tripartite mutual information and entanglement entropy vs interaction range parameter $\alpha$ for two values of circuit density $M_2/N$ and $N$ ranging from 64 (lightest curves) to 512 (darkest curves). Entropy values are weighted by $\log N$ to highlight the regions of critical entanglement growth for $M_2/N = 0.2, 0.5$, where the curves collapse on each other. The curves of $\bar{\mathcal{I}}_3$ similarly appear to collapse on each other in this region, with values near zero but positive.}
\label{fig:sb-crossover}
\end{figure}

We examine the trends of entanglement growth by comparing linear and logarithmic fits of $\bar{S}$ versus system size $N$ (lower left panel of Fig. \ref{fig:sb-pd}). Two regimes emerge: in long-range measurement circuit, entanglement grows linearly with system size, consistent with volume-law entanglement, while in the short-range measurement circuits, entanglement grows logarithmically, consistent with critical and sub-volume law entanglement growth. Fig. \ref{subfig:sb-sfit} examines the entanglement growth from four points across the phase diagram, confirming the linear growth for long-range measurement circuits and logarithmic growth for short-range measurement circuits. In the right panels of Fig. \ref{fig:sb-crossover}, the steady state entropy normalized by $\log N$ is plotted versus $\alpha$ for different system sizes. The curves collapse on each other where the measurements are short-range, which is consistent with the observed logarithmic growth of $\bar{S}$ with $N$.

These results demonstrate the presence of an entanglement transition in the measurement range $\alpha$ for all circuit densities. The transition shifts to favor logarithmic entanglement as the circuit density increases, occurring near $\alpha = 3$ when $M_2 / N \rightarrow 0$ and near $\alpha = 2$ when $M_2 / N = 0.5$. We suspect that the universality class of the transition remains the same with the random-basis MoC as the circuit density increases, albeit the transition point shifts to favor sub-volume law entanglement. Since the circuit is more structured due to the extensive set of commuting measurements within a layer, the direction of this shift is not surprising. Therefore, the formalism developed for random-basis MoC based on Haar unitaries can still be useful to predict the universal physics of MoC with power-law decaying interactions, where the choice of single-basis within a layer acts as a microscopic detail. 

Quantitatively comparing the data to the random-basis case in Fig. \ref{subfig:rb-sfit}, we note that the system entanglement is smaller for the single-basis case, particularly as circuit density is increased. This is expected as the circuit now possesses more structure. Particularly a single-basis dense circuit prepares an exceptionally disordered state, causing the temporal boundary conditions to propagate only a few layers into the system before the state is fully purified (see purification data in Fig.~\ref{subfig:sb-taufit}). Entanglement entropy, calculated from the free energy cost of changing the final time boundary condition in the statistical mechanics model, is therefore significantly lower than in the random-basis case.

\subsubsection{Purification}\label{sec:sb-purification}
In the lower right plot of Fig. \ref{fig:sb-pd}, we probe the purifying capacity of single-basis MoCs. This plot visualizes the difference in $R^2$ values when the $\tau$ versus $N$ data is fit to an exponential (non-purifying) and a linear (purifying) function. The sparsest ($M_2/N \sim 0$) and long-range measurement regime is found to have $\tau(N) \sim e^N$ -- as it must, in order to match the random-basis results in the limit $M_2/N \rightarrow 0$. However, as the circuit density increases, the purification becomes possible across all measurement ranges. 

The densest limit of the circuit, $M_2/N=0.5$ is omitted from this particular phase diagram (depicted with a gray line) due to the fact that neither $Ae^{kN}$ nor $aN + b$ properly describes the observed $\tau(N)$ data. However, at the top row of Fig. \ref{subfig:sb-taufit}, we plot $\tau(N)$ for $\alpha = 0$ (left), $ = 4$ (right) and find that the purification timescale is independent of system size $N$. Moreover, purification occurs exceptionally quickly, with $\tau \approx 1$ for the $M_2/N = 0.5$ single-basis MoCs. Away from this limit (bottom row of Fig. \ref{subfig:sb-taufit}) for rather sparse circuits, linear growth of $\tau$ with $N$ is observed. 

Hence, we observe that the purification data is insensitive to the entanglement transition in the measurement range for most of circuit densities. In other words, in a dense and long-range circuit, volume-law entanglement is surprisingly accompanied with a $\mathcal{O}(1)$ purification timescale, and this quick purification occurs regardless of the entanglement entropy scaling. We can understand this subtlety by once more invoking the idea of \textit{measurement frustration} \cite{PhysRevX.11.011030}. In addition to considering how likely it is for individual pairs of measurement observables to commute with each other, here we find that we must also consider how likely it is for \textit{products} of measurement observables to commute with other products of measurement observables. In the stabilizer picture, once a set of N independent mutually commuting parity checks (or products of parity checks) have been accumulated, the state purifies, i.e.,~it becomes independent of its starting conditions. In the single-basis design, this happens quicker as the density increases. Consider, for instance, a dense circuit where a layer of $Z$-basis measurements is followed by a layer of X-basis measurements. If \(X_iX_j\) is measured in the X-basis layer, it will not necessarily commute with the two body stabilizers from the previous layer, say \(Z_kZ_i\) and \(Z_jZ_l\), but it will always commute with their product $Z_iZ_jZ_kZ_l$, which is also a stabilizer. Thus $X_iX_j$ is added to the stabilizer group while the string $Z_iZ_jZ_kZ_l$ is preserved because it commutes with $X_iX_j$. More generally, two Pauli operators commute whenever the number of sites on which they act with different non-identity Paulis is even. As the density increases, the circuit rapidly builds a large set of mutually commuting stabilizers, and an increasing fraction of subsequent measurements either commute with (or are redundant given) the existing stabilizer group. This reduces measurement frustration and makes many outcomes effectively deterministic, thereby accelerating purification even when the conditional state of the system might retain volume-law entanglement.

Additionally, in the single-basis protocol there is a $1/3$ probability that two consecutive layers are performed in the same Pauli basis. Such repeated-basis layers are fully commuting and therefore act as comparatively strong projectors onto a common set of stabilizer constraints. This effect is most pronounced in the dense limit $M_2/N \rightarrow 0.5$, where a single repeated-basis layer supplies $O(N)$ commuting parity checks and can rapidly build a large mutually compatible stabilizer algebra, leading to fast purification and a suppression of information scrambling, as observed in Fig.~\ref{fig:sb-pd}.

\subsubsection{Scrambling and Non-Scrambling Regimes}

The steady state behavior of $\mathcal{I}_3$ is quite nontrivial across the parameter space for the single-basis circuit design, (top left corner in Fig.~\ref{fig:sb-pd}). For sparse circuits, the value of $\bar{\mathcal{I}}_3$ transitions from negative values with large magnitude in the long-range circuit to small but positive values in short-range circuits. Hence, we observe that the TMI tracks closely the entanglement transition in the measurement range. Fig.~\ref{fig:sb-crossover} left-bottom panel shows the transition from scrambling, $\bar{\mathcal{I}}_3 \ll 0$, to non-scrambling, $\bar{\mathcal{I}}_3 \sim 0$, regime. This change from the random-basis circuits, where a transition was not conclusively observed, occurs due to a density-driven scrambling transition that becomes visible especially at short range. 

As circuit density increases, the scrambling regime shrinks in size, completely disappearing for very dense circuits. 
The overall reduction in scrambling due to increased circuit density can be attributed to the fact that the stabilizers of the state become long single-basis Pauli strings and are thus more likely to commute with the following circuit layers, as discussed in the extended measurement frustration argument in Section \ref{sec:sb-purification}. 

Away from the extreme-density limits, consecutive layers typically alternate between noncommuting bases and the measurement range $\alpha$ again controls how efficiently nonlocal constraints are generated and propagated. Consistent with this, we observe a scrambling-to-nonscrambling transition as a function of $\alpha$ across a wide region of the single-basis phase diagram, with the densest and sparsest regions providing notable exceptions where the layerwise commuting-projector effect (dense) or the effective equivalence to the random-basis protocol (sparse) dominates.

Finally, due to the disappearance of scrambling transition in the measurement range in densest circuit designs, a non-scrambling phase with volume-law entanglement entropy emerges at $\alpha \lesssim 2$. Next, we discuss this novel phase in more detail in the following section.

\subsubsection{Mutual information and a regime of Bell-pairs}\label{bell pair section}

In single-basis MoC, there is considerably more mutual information, and hence long-range entanglement, in the generated steady states compared to the random-basis design, see the comparison in Fig.~\ref{subfig:Ivsr}. We find Bell pairs across the entire chain when the circuit is dense, regardless of the measurement range. The distribution of Bell pairs differs however with uniform in the long-range, and a skewed distribution peaking at the shortest distance, $r=1$, when the circuit is short-range. The spatial profile of mutual information mirrors these distinct behaviors in the Bell pair distributions: we observe a constant profile in the long-range circuit and a slow decay to a constant in the short-range. 

The most prominent regime with significant mutual information is the dense and long-range region (upper right panel in Fig. \ref{fig:sb-pd}). These states remarkably exhibit volume-law entanglement with no scrambling. However due to the presence of volume-law rather than area-law, they differ from the states generated in the projective XXZ model that also had large mutual information with no scrambling. 

The entanglement transition in the sparse single-basis circuits is also detected by the mutual information, which exhibits a constant spatial profile and a power-law spatial decay, in long- and short-range circuits, respectively. Correspondingly, the Bell pair distribution is uniform and Poisson-like in these competing phases, suggesting the presence and absence of long-range entanglement.

\section{Discussion and Outlook}\label{Sec3}

Measurement-only circuits have surprisingly rich entanglement phases that require observables beyond the entanglement entropy for a classification. We achieved this classification by further examining mutual information, tripartite mutual information, purification dynamics and Bell pair counting. We systematically expanded the investigation initiated by Ref.~\cite{kuno_phase_2023} in power-law decaying measurement circuits. This expansion was performed in a multi-faceted way: 

A central theoretical contribution of this work is the mapping of the (trajectory-averaged) entanglement entropy in 1+1D circuits to free energy cost in a 2D statistical mechanics model by developing a formalism for random-basis MoC. This formalism was then used to show the emergence of an effective XX hamiltonian in the continuous time limit of the imaginary time evolution. The developed method predicted a volume-law to sub-volume law entanglement transition at range $\alpha \sim 3$ by revealing the mapping to continuous symmetry breaking (CSB) and critical XY phases, respectively, in the effective Hamiltonian. We confirmed this theoretical prediction with large scale Clifford simulations revealing a measurement range driven entanglement transition for most circuit densities. Intriguingly, the transition in entanglement entropy persists when more structure is introduced to the circuit, i.e.,~single-basis design, although the transition point shifts as the circuit density increases. Due to smoothness of this shift, we argue that the entanglement transition in the single-basis MoC must be governed by the same universal properties of the transition in the random-basis MoC.

We introduce \textit{the measurement density} to the circuit design as a new circuit parameter and showed the presence of entanglement and scrambling transitions across this new parameter. An entanglement transition occurred in short-range and random-basis MoC, as well as in single-basis MoC in the interval of ranges $\alpha \sim (2,3)$. 
In the random-basis protocol, tripartite mutual information changes in tandem with the entanglement transition, but our steady-state finite-size analysis does not support a definitive claim of a scrambling transition, in contrast to some interpretations of Ref.~\cite{kuno_phase_2023}. In the single-basis protocol, by contrast, we observe clear scrambling transitions driven by both measurement range and density over broad regions of parameter space.
Furthermore, our work demonstrates that the random-basis MoC exhibits scrambling where the speed of scrambling is in fact set by the density. Contrary to Ref.~\cite{kuno_phase_2023}, the sparse circuit limit exhibits scrambling at a speed of $N \log N$ in its steady state. As the circuit becomes denser, genuine fast scrambling onsets with $\log N$. 

The symmetry breaking phase of the effective Hamiltonian in random-basis design is accompanied by negligible mutual information in the circuit steady state, even though Bell pairs at all ranges were detected. This highlights that long-range order in the effective Hamiltonian description does not necessarily imply long-range entanglement in the corresponding circuit steady state \cite{PhysRevA.93.012303}.

We further constructed a projective XXZ model in sparse and short-range MoC by varying the probability of measuring in $Z$-basis. Intriguingly, this revealed a sub-volume to area-law entanglement transition that was accompanied by a transition in the mutual information from power-law decay to constant function in distance, exactly at the SU(2) symmetric point. While our effective XX hamiltonian description captures the symmetric point, complete analytical theory for the full anisotropic model, in particular for why the \(Z\)-dominated regime exhibits Ising phase-like features while the \(X/Y\) dominated regime appears XY phase-like, remains an open problem.

Our results show that measurement-only circuits can stabilize exotic steady state phases through the interplay of measurement range and density. For example, random-basis design features a volume-law entangled, scrambling and non-purifying phase with  power-law decaying mutual information in distance, raising the question of why power-law correlations persist away from the criticality. 
In the single-basis protocol, multiple distinct regimes emerge, including phases where entanglement and purification are no longer tightly linked: volume-law entanglement can coexist with purification, and in certain parameter regimes scrambling properties change while entanglement scaling remains the same.
Finally, and notably, a volume-law entangled phase is found to purify rapidly, with an $O(1)$ purification timescale that shows no discernible dependence on system size. This phase also does not scramble information and exhibits long-range entanglement, suggesting an efficient measurement-only route to preparing highly entangled resource states that may be technologically useful.

\section{Acknowledgments} 

We thank for stimulating discussions with Yidan Wang, Andi Gu, Sarang Gopalakrishnan, Soonwon Choi.  
A.M.G acknowledge support from the NSF through the Graduate
Research Fellowships Program, and A.M.G also acknowledges support
through the Theodore H. Ashford Fellowships in the Sciences. F.A.-M. acknowledges support from an MIT first year fellowship, and F.A.-M. also acknowledges support from Harvard through the Program for Research in Science and Engineering (PRISE).
H.-Y.H. and S.F.Y. acknowledge funding from the NSF through the CUA PFC (PHY-2317134) and the DOE QUACQ grant (DE-SC0025572).
C.B.D acknowledges support from the NSF through a grant for ITAMP at Harvard University. 

\bibliographystyle{apsrev4-1}
\bibliography{Bibliography} 

\appendix
\section{Simulation Details \label{appendix:sim}}

\begin{figure*}[!htb]
\centering
\subfloat[][]{
\includegraphics[width=1.2\columnwidth]{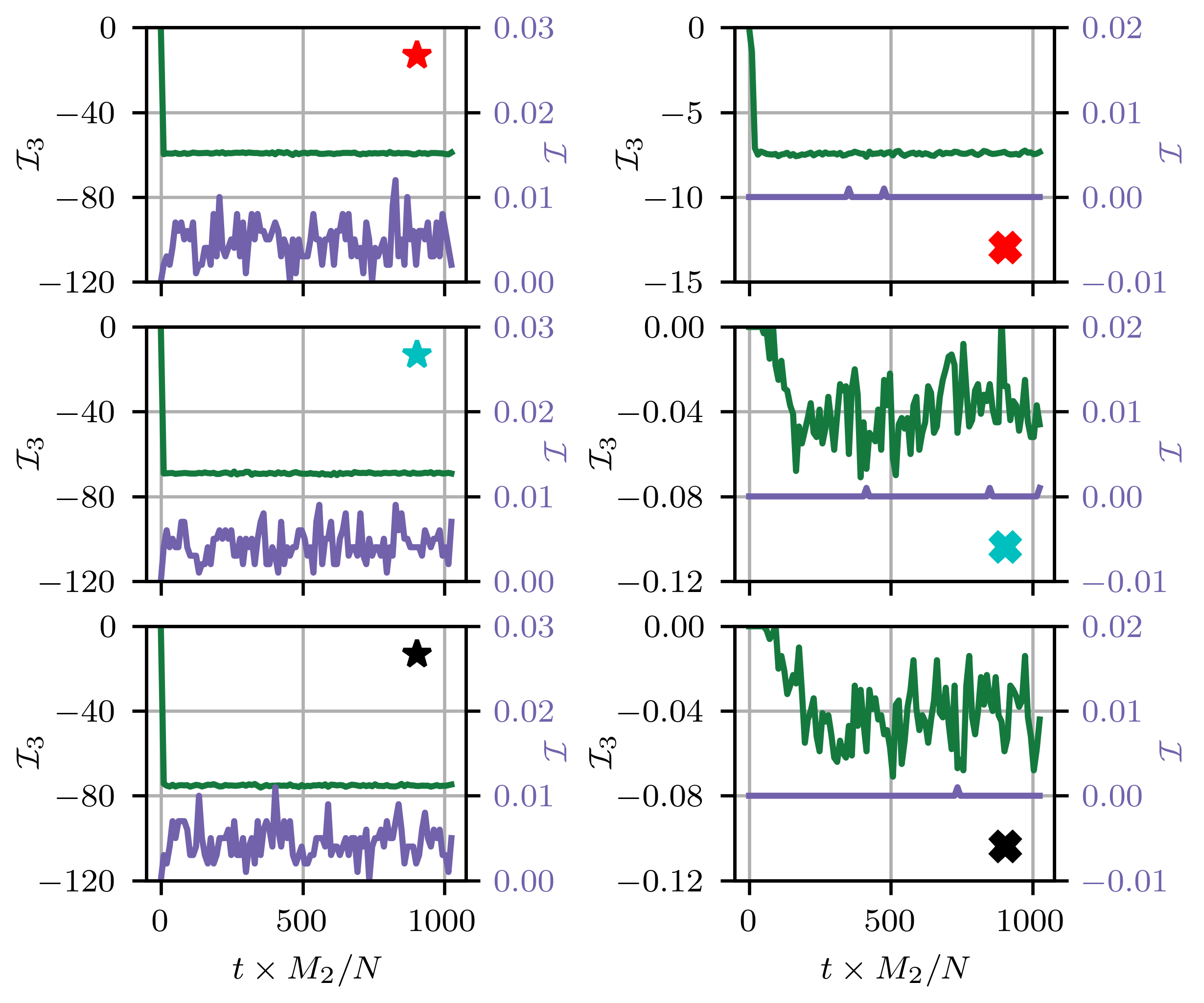}\label{subfig:rb-II3}
}
\hfill
\subfloat[][]{
\includegraphics[width=0.8\columnwidth]{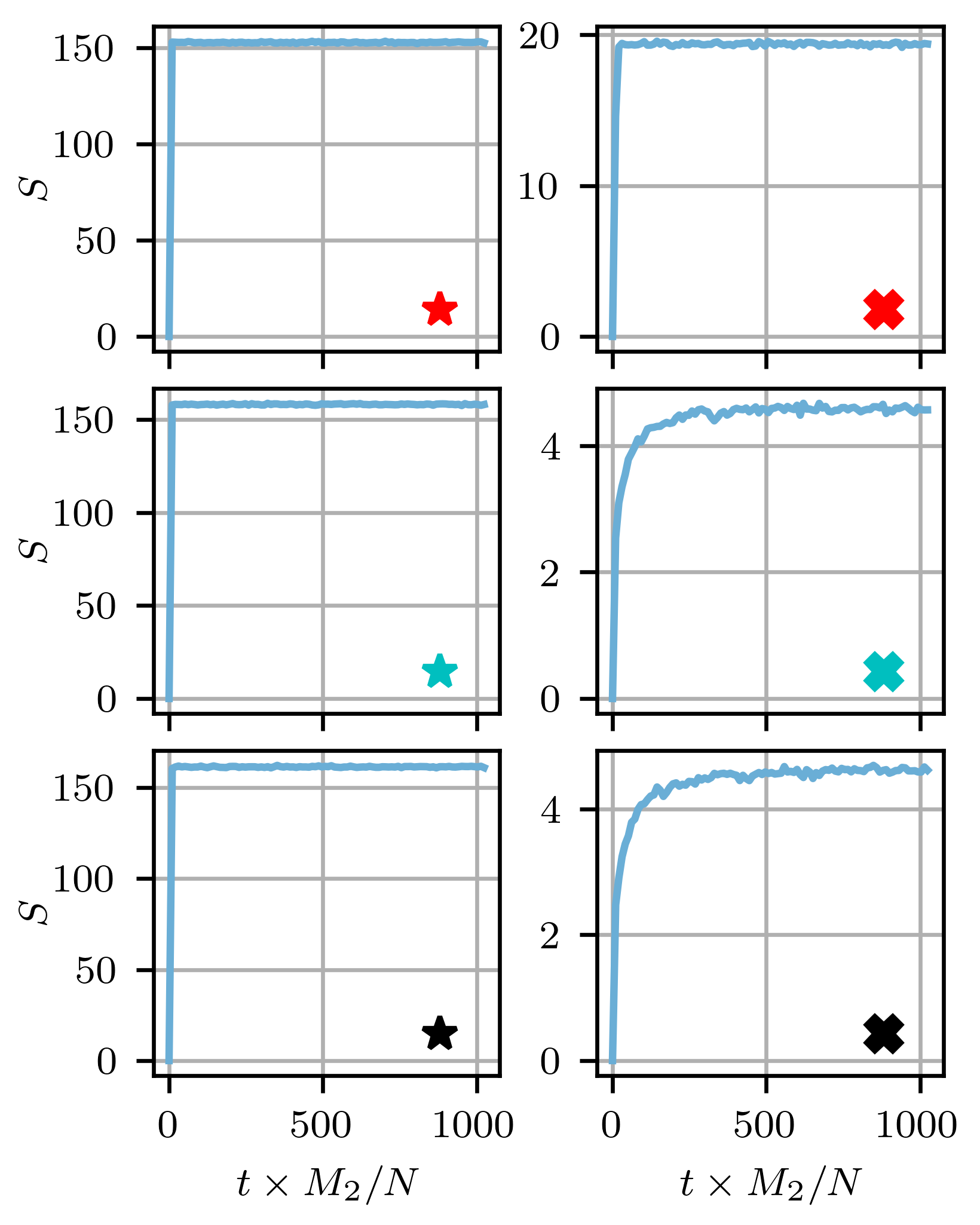}\label{subfig:rb-S}
}
\caption{Observed values of $\mathcal{I}_3$, $\mathcal{I}$ (\ref{subfig:rb-II3}), and $S$ (\ref{subfig:rb-S}) for a single circuit trajectory at six points across the random-basis phase diagram, as indicated by the matching marker and marker color in Fig. \ref{fig:rb-pd}. Left column: long-range measurements ($\alpha = 0$); right column: short-range measurements ($\alpha = 4$); from bottom to top: circuit density increasing with values $M_2/N = 0.0, 0.2, 0.5$. Here, $N = 512$.}
\label{fig:rb-traj}
\end{figure*}

\begin{figure*}[!htb]
\centering
\subfloat[][]{
\includegraphics[width=1.2\columnwidth]{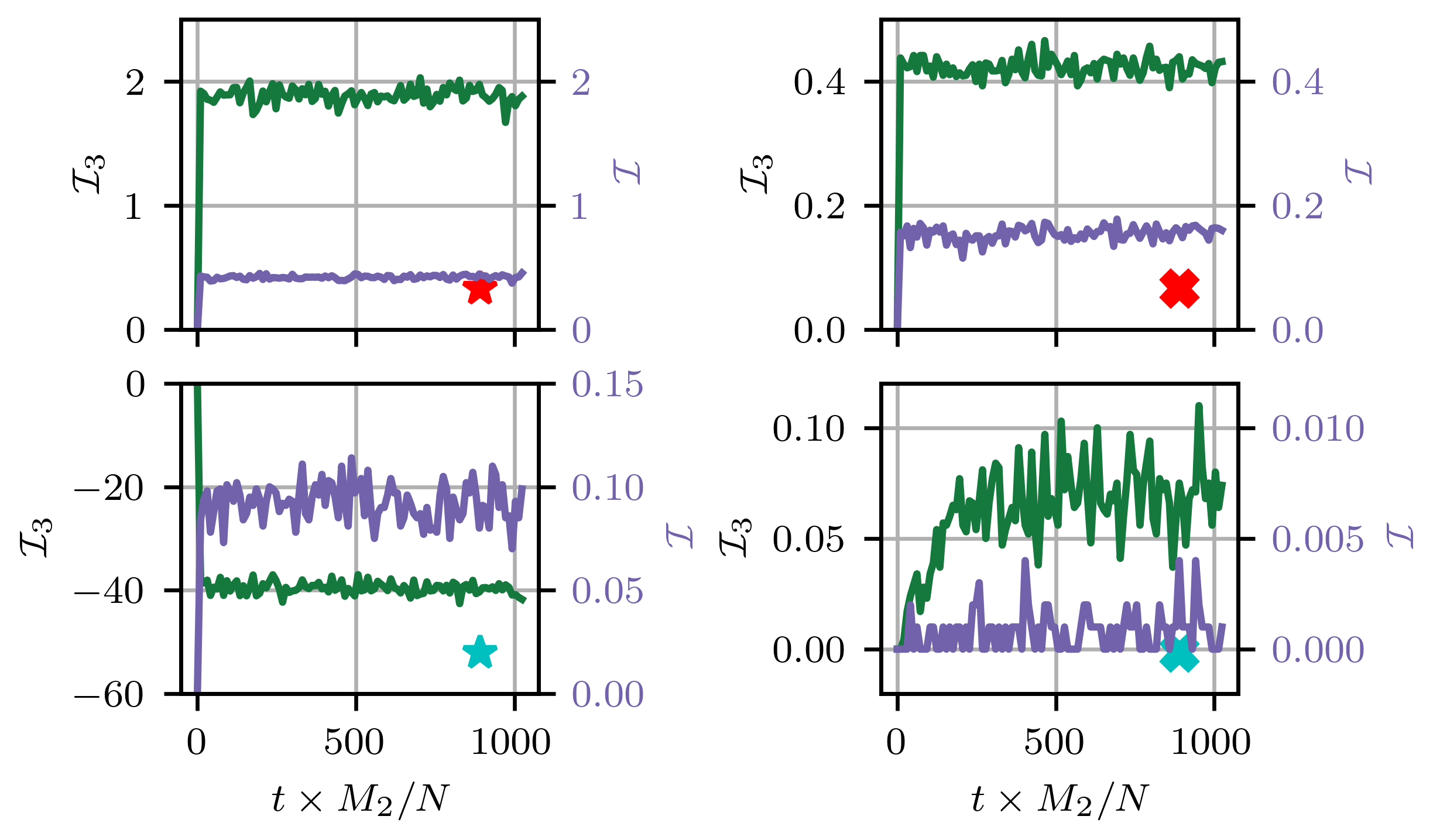}
\label{subfig:sb-II3}
}
\hfill
\subfloat[][]{
\includegraphics[width=0.8\columnwidth]{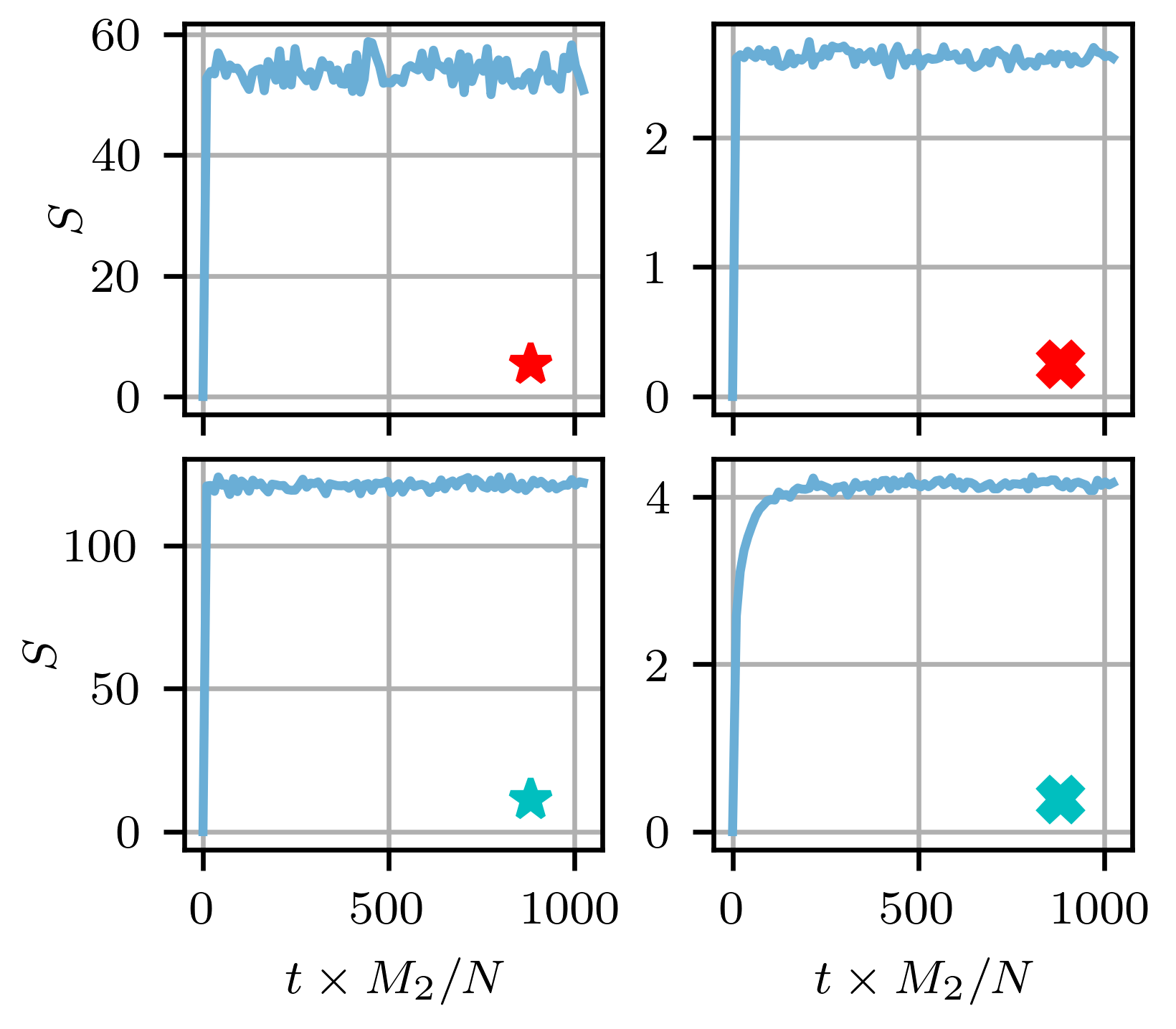}
\label{subfig:sb-S}
}
\caption{Observed values of $\mathcal{I}_3$, $\mathcal{I}$ (\ref{subfig:sb-II3}), and $S$ (\ref{subfig:sb-S}) for a single circuit trajectory at four points across the single-basis phase diagram, as indicated by the matching marker and marker color in Fig. \ref{fig:sb-pd}. Left column: long-range measurements ($\alpha = 0$); right column: short-range measurements ($\alpha = 4$); from bottom to top: circuit density increasing with values $M_2/N = 0.2, 0.5$. Here, $N = 512$.}
\label{fig:sb-traj}
\end{figure*}

We utilize the package PyClifford to perform our Clifford circuit simulations. In this appendix, we provide a few examples of individual circuit trajectories to demonstrate the depth of our simulation (Figs. \ref{fig:rb-traj}, \ref{fig:sb-traj}). In order to establish steady state, we simulate circuits of a depth $2N^2 / M_2$ (up to 524,288 layers for the largest system considered, $N = 512$). We save computational resources by only computing observables at 100 intermediate points during the evolution. For timescale estimation (that is, computing time to scrambling and purification time), we perform simulations at higher temporal resolution over a smaller region of $t$. 

\begin{figure}[!htb]
\centering
\includegraphics[width=1.0\columnwidth]{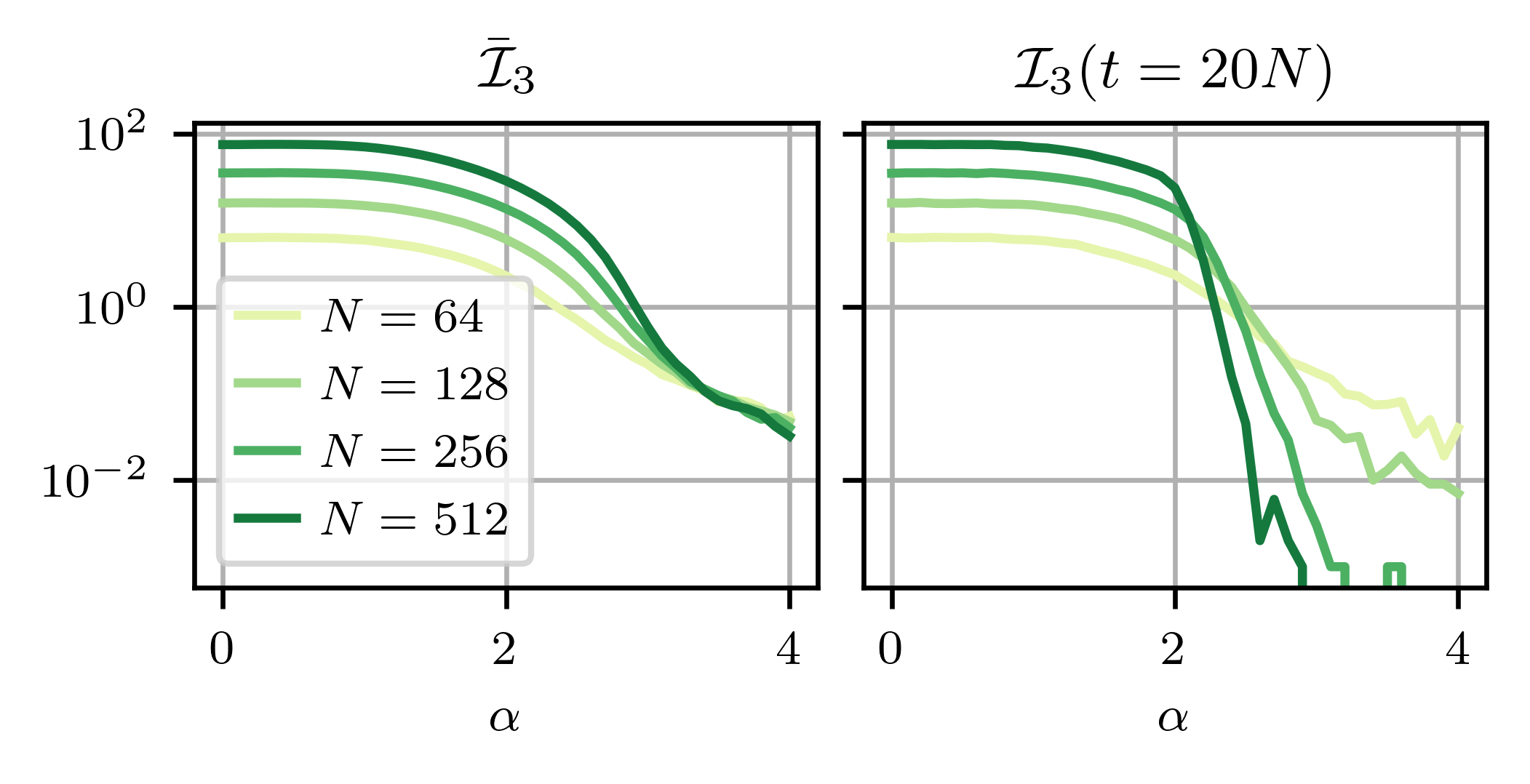}
\caption{The effect of simulation depth. In the left panel, we plot the steady state value of $\mathcal{I}_3$, as defined in the main text. In the right panel, we plot the trajectory-averaged value of $\mathcal{I}_3$ at a depth $t = 20 N$. At this point in the simulation, the trajectory-averaged $\mathcal{I}_3$ has not reached steady state, particularly simulations of large system size $N$ and large $\alpha$. This results in an apparent crossover in $\mathcal{I}_3$.}
\label{fig:paper-comp}
\end{figure}

If simulations are not sufficiently long, particularly as system size is increased, there is a risk of fabricating system size dependence that is actually a byproduct of short-time simulations. To illustrate the importance of long-time simulations in establishing steady state for our system, we compare our results to those of Ref. \cite{kuno_phase_2023}. In the limit of $M_2/N = 0$, our system is identical to their circuit setup except for a difference in the definition of time (in \cite{kuno_phase_2023}, a factor of $N$ is implicitly included in their definition of one time step). The authors perform simulations up to $t = 20 N$ (converting to our definition of $t$) and claim to detect a crossover in $\mathcal{I}_3$ at a critical value of $\alpha$. We reproduce their results in the right panel of Fig. \ref{fig:paper-comp} by performing shorter depth simulations and indeed observe what appears to be a crossover at a value of $\alpha$ between 2 and 3. However, if we create the same plot with simulations of larger depth (left panel of Fig. \ref{fig:paper-comp}), no crossover is observed. If a true scrambling transition exists, it significantly shifts relative to the claim of \cite{kuno_phase_2023} to considerably favor the scrambling regime. 

\begin{figure*}[!htb]
\centering
\subfloat[][]{
\includegraphics[width=1.0\columnwidth]{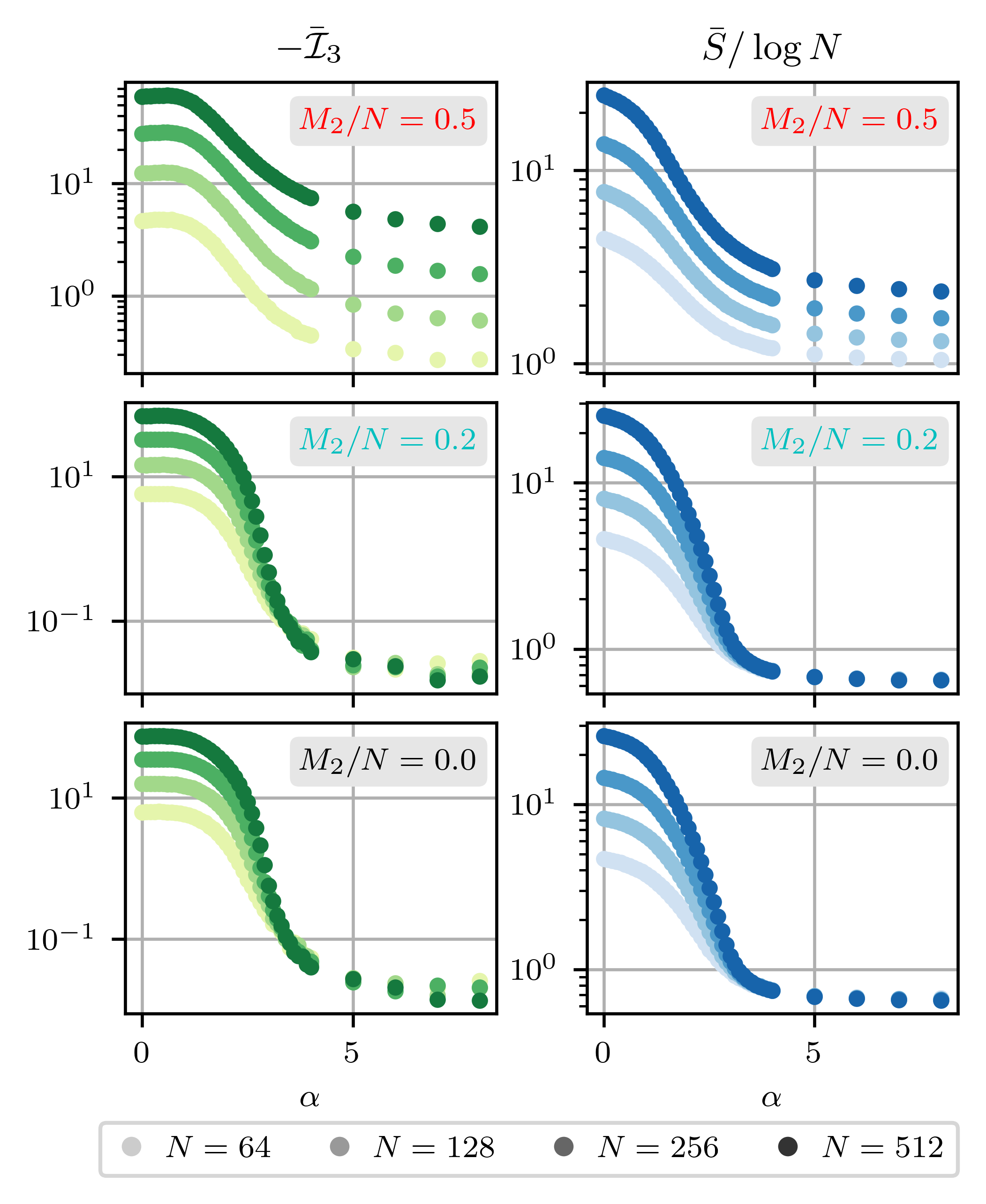}\label{subfig:rb-LogCrossover}
}
\hfill
\subfloat[][]{
\includegraphics[width=1.0\columnwidth]{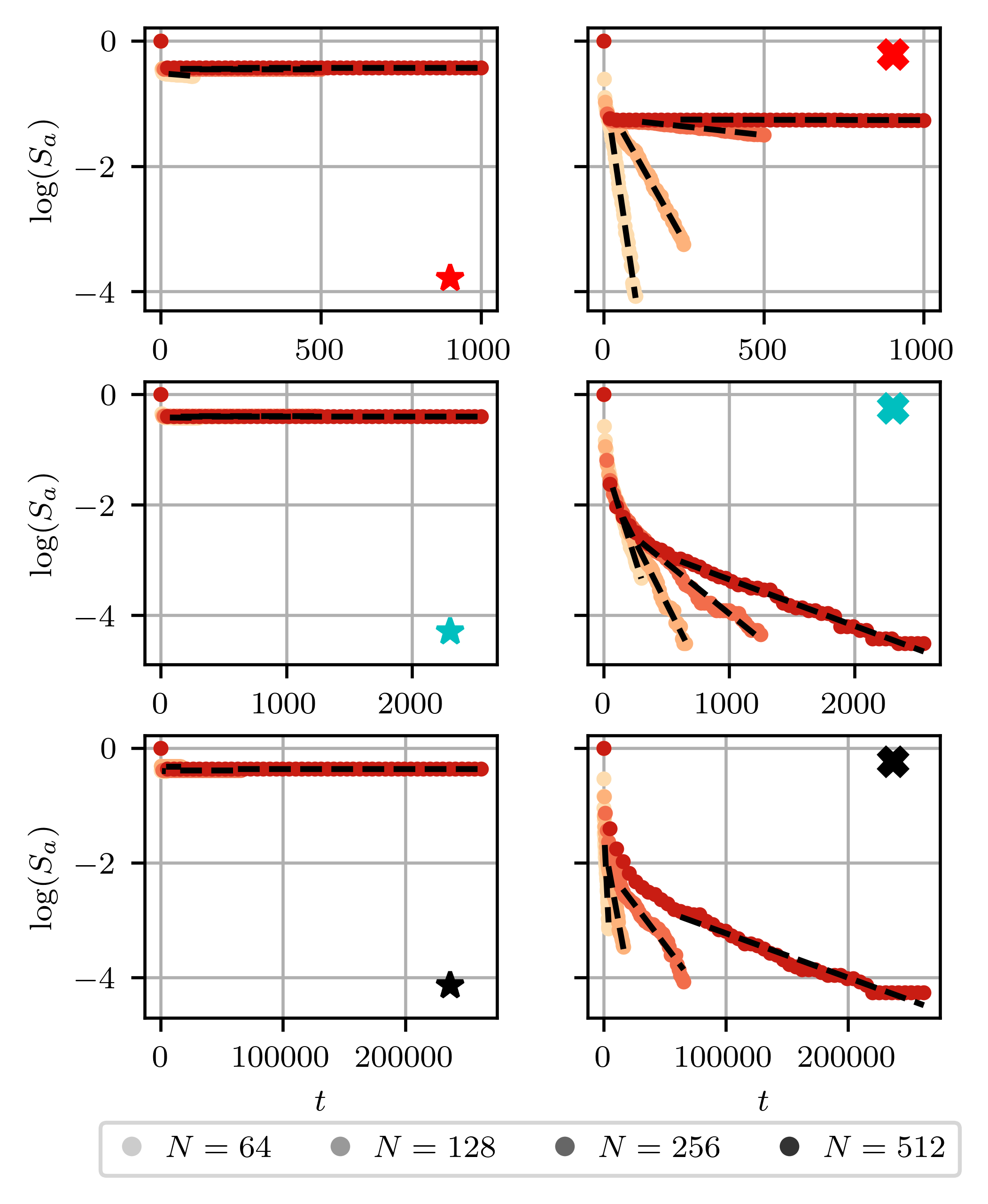}
\label{subfig:rb-SaScaling}
}
\caption{\ref{subfig:rb-LogCrossover}: Steady state values of tripartite mutual information and entanglement entropy vs interaction range parameter $\alpha$ for three values of circuit density $M_2/N$ and $N$ ranging from 64 (lightest curves) to 512 (darkest curves). Entropy values are weighted by $\log N$ to highlight the regions of critical entanglement growth for $M_2/N = 0.0, 0.2$, where the curves collapse on each other. The curves of $\bar{\mathcal{I}}_3$ similarly appear to collapse on each other in this region, with values near zero. For the densest circuits (top row), no collapse is observed. As opposed to Fig. \ref{fig:rb-crossover} in the main text, here we examine the results with a log scale on the y-axis with a few additional values of $\alpha$ far in the short-range regime. The results confirm that no true crossover in TMI is observed up to $\alpha = 8.0$. \ref{subfig:rb-SaScaling}: Scaling of ancilla entanglement entropy $S_a$ with $t$ for six points across the random-basis phase diagram, as indicated by matching marker and marker color in Fig. \ref{fig:rb-pd}. The log of $S_a$ is taken to highlight the exponential decay of $S_a$ with $t$, which is particularly clear in the right column (short-range measurements). The values of $\tau$ are extracted from this data by fitting the exponential decay of these curves ($S_a \sim e^{-t/\tau}$), as indicated by the black dashed lines.}
\label{fig:rb-LogCrossover-SaScaling}
\end{figure*}

\begin{figure*}[!htb]
\centering
\subfloat[][]{
\includegraphics[width=1.0\columnwidth]{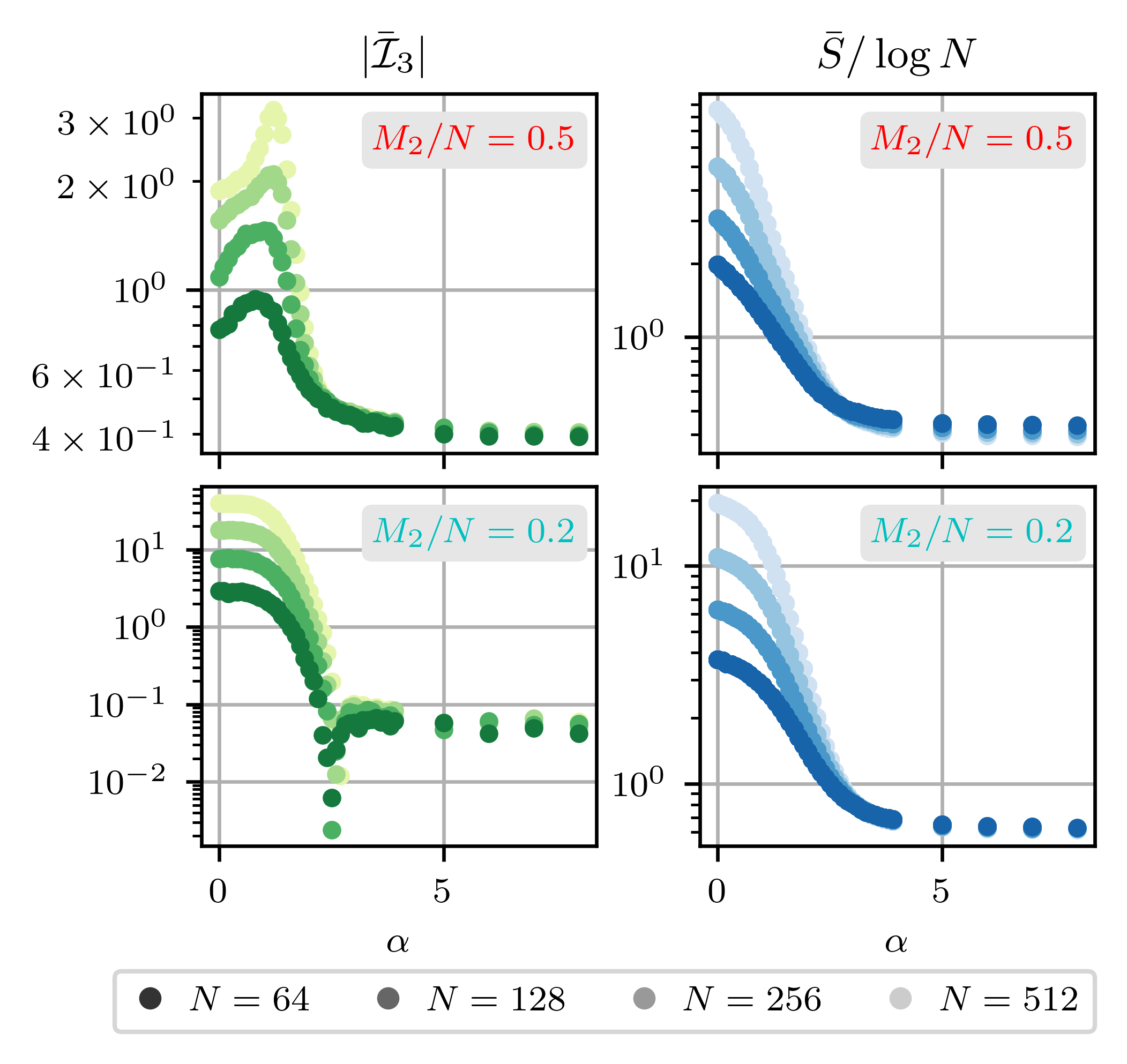}\label{subfig:sb-LogCrossover}
}
\hfill
\subfloat[][]{
\includegraphics[width=1.0\columnwidth]{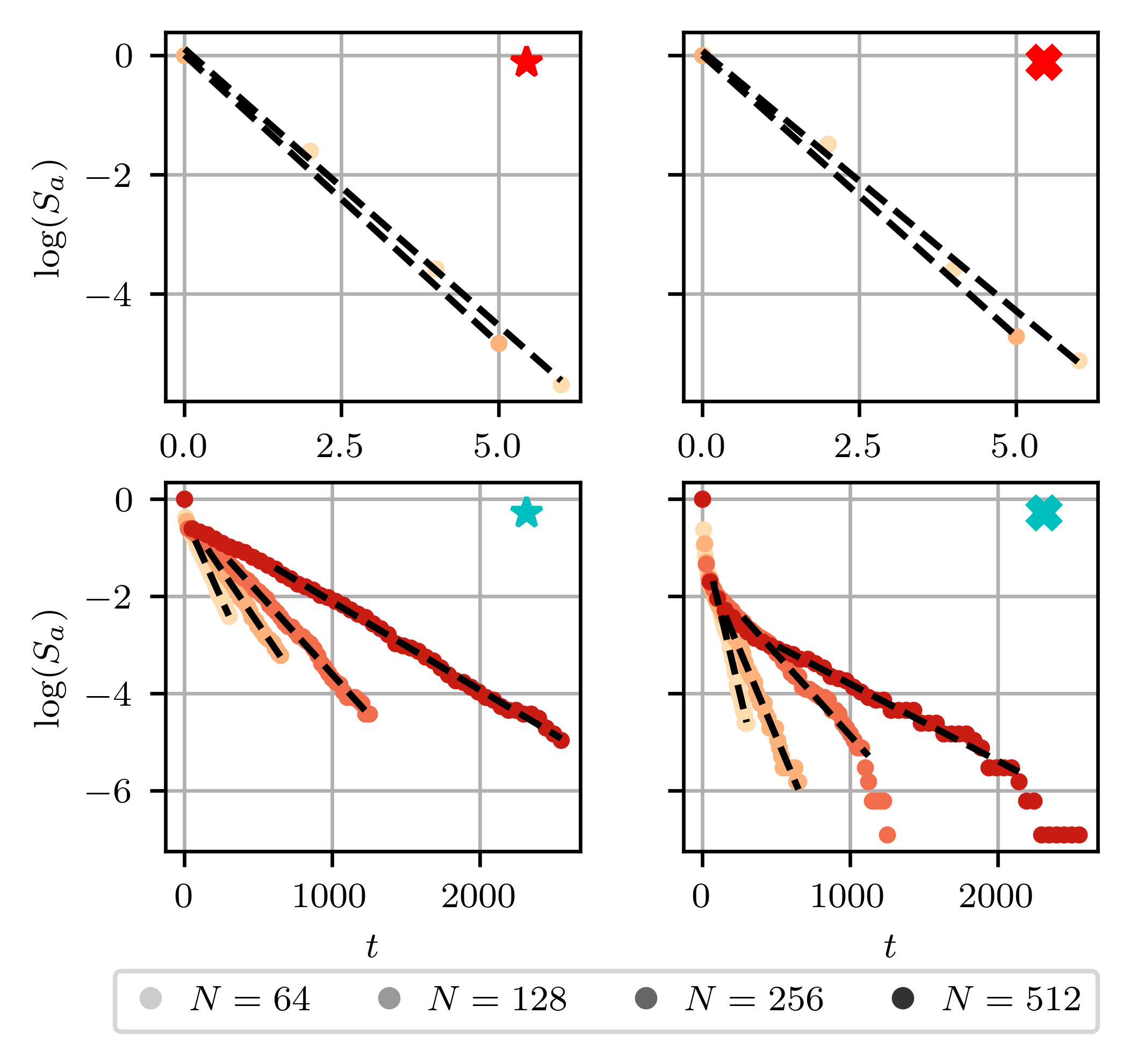}\label{subfig:sb-SaScaling}
}
\caption{\ref{subfig:sb-LogCrossover}: Steady state values of tripartite mutual information and entanglement entropy vs interaction range parameter $\alpha$ for two values of circuit density $M_2/N$ and $N$ ranging from 64 (lightest curves) to 512 (darkest curves). Entropy values are weighted by $\log N$ to highlight the regions of critical entanglement growth for $M_2/N = 0.2, 0.5$, where the curves collapse on each other. The curves of $\bar{\mathcal{I}}_3$ similarly appear to collapse on each other in this region, with values near zero but positive. As opposed to Fig. \ref{fig:sb-crossover} in the main text, here we examine the results with a log scale on the y-axis with a few additional values of $\alpha$ far in the short-range regime. The results confirm that no true crossover in TMI is observed up to $\alpha = 8.0$. \ref{subfig:sb-SaScaling}: Scaling of ancilla entanglement entropy $S_a$ with $t$ for four points across the single-basis phase diagram, as indicated by matching marker and marker color in Fig. \ref{fig:sb-pd}. The log of $S_a$ is taken to highlight the exponential decay of $S_a$ with $t$. The values of $\tau$ are extracted from this data by fitting the exponential decay of these curves ($S_a \sim e^{-t/\tau}$), as indicated by the black dashed lines.}
\label{fig:sb-LogCrossover-SaScaling}
\end{figure*}
In Figs. \ref{subfig:rb-LogCrossover} and \ref{subfig:sb-LogCrossover}, we take a second look at the TMI and entanglement entropy of the random-basis and single-basis MoCs, respectively. We include additional data at $\alpha = 5, 6, 7, 8$ and use a log-scale to see more subtle changes in the data in the short-range measurement regime, yet see no change in the previous trends and no crossovers are observed. 

Finally, in Figs. \ref{subfig:rb-SaScaling} and \ref{subfig:sb-SaScaling}, we provide the scaling of the ancilla qubit entanglement entropy $S_a$ with time $t$. The purification timescale $\tau$ is computing by fitting the exponential decay of $S_a$ with $t$.

\section{Scalings of cut-crossing measurements}\label{bipartition_scaling}

Here we estimate how the expected number of two-qubit measurements that \textit{cross a fixed bipartition cut} scales with system size \(N\) and changes with the measurement-range exponent \(\alpha\) and the number of measurements per layer \(M_2\). This provides a simple geometric proxy for how readily bipartite entanglement across a cut can be generated on the simulation time window \(t\sim N\) of random-basis MoC. 

Considering a 1D ring of \(N\) qubits with a fixed cut, a two-qubit measurement acts on a pair of sites separated by a distance \(r\in\{1,\dots,N/2\}\), where the separation is sampled from a power-law distribution
\begin{equation}
    P(r)=\frac{r^{-\alpha}}{Z_N},\qquad 
    Z_N:=\sum_{r=1}^{N/2} r^{-\alpha}.
\end{equation}
Approximating sums by integrals (up to \(N\)-independent corrections), the mean separation scales as
\begin{equation}
    \mathbb{E}[r] = \int_1^{N/2} rP(r) dr\sim
    \begin{cases}
        N, & \alpha<1,\\[3pt]
        \dfrac{N}{\log N}, & \alpha=1,\\[8pt]
        N^{2-\alpha}, & 1<\alpha<2,\\[3pt]
        \log N, & \alpha=2,\\[3pt]
        \text{const.}, & \alpha>2.
    \end{cases}
    \label{eq:Er_scaling_app}
\end{equation}
To relate \(\mathbb{E}[r]\) to cut crossings, note that for a fixed cut the probability that a randomly placed bond of length \(r\) straddles the cut is proportional to \(r/N\). Concretely, if one first chooses a site \(i\) uniformly and then chooses its partner at separation \(r\), the event that the pair \((i,i+r)\) crosses the cut occurs with probability
\begin{equation}
    \Pr(\text{cross}\mid r)\sim \frac{r}{N}.
\end{equation}
Therefore, the cut-crossing probability for a single measurement scales as
\begin{equation}
    \mathbb{E}[\mathbf{1}_{\text{cross}}]
    =\sum_{r}P(r)\Pr(\text{cross}\mid r)
    \sim \frac{1}{N}\,\mathbb{E}[r].
    \label{eq:cross_per_meas_app}
\end{equation}
If a circuit layer contains \(M_2\) measurements, the expected number of cut-crossing measurements per layer is then
\begin{equation}
    \mathbb{E}[\#\text{crossings per layer}]
    \sim M_2\,\frac{\mathbb{E}[r]}{N}.
    \label{eq:cross_per_layer_app}
\end{equation}

In our simulations we evolve to depths proportional to system size, \(t\sim N\). Over this window, the accumulated number of cut-crossing measurements scales as $M_2\,\mathbb{E}[r]$.

Two limits are particularly transparent. In the asymptotically sparse case \(M_2=O(1)\), so that \(M_2/N\to 0\)), total number of crossings over time scales as \(\sim\mathbb{E}[r]\), i.e., it is \(O(1)\) for \(\alpha>2\), logarithmic at \(\alpha=2\), subextensive \(N^{2-\alpha}\) for \(1<\alpha<2\), and extensive for \(\alpha\le 1\) (with a \(N/\log N\) correction at \(\alpha=1\)), as summarized in Eq.~\eqref{eq:Er_scaling_app}. In the dense case of fixed measurement density \(\rho=M_2/N\), instead we have \(\mathbb{E}[\#\text{crossings up to }t\sim N]\sim \rho N\,\mathbb{E}[r]\), which is \textit{at least} \(O(N)\) for all \(\alpha\) and grows faster for smaller \(\alpha\).

In the statistical mechanics mapping used in the main text, interaction blocks that straddle the identity-swap interface are precisely those that can contribute an \(O(1)\) free-energy penalty to that interface; thus the scaling estimates above provide a geometric criterion for how many such interface-straddling blocks are available on the \(t\sim N\) time window.

\end{document}